\documentclass[preprint]{aastex}
\usepackage{emulateapj5, psfig, epsfig}
\slugcomment{accepted by The Astronomical Journal}

\shorttitle{The Draco and UMi dwarf spheroidals}
\shortauthors{Bellazzini, Ferraro \& Pancino}

\begin{document}

\title{The Draco and Ursa Minor dwarf spheroidals. \\ A comparative study.
    \thanks{Based on observations made with the Italian Telescopio Nazionale
     Galileo (TNG) operated on the island of La Palma by the Centro
  Galileo Galilei of the CNAA (Consorzio Nazionale per l'Astronomia e
 l'Astrofisica) at the Spanish Observatorio del Roque de los Muchachos
                  of the Instituto de Astrofisica de Canarias.}}

\author{M. Bellazzini, F.R. Ferraro, L. Origlia}
\affil{INAF - Osservatorio Astronomico di Bologna, Via Ranzani 1, 40127, 
Bologna, ITALY}
\email{bellazzini,ferraro,origlia@bo.astro.it}

\author{E. Pancino\thanks{Dipartimento di Astronomia - 
        Universit\`a di Bologna, via Ranzani 1, 40127, Bologna, ITALY}}
\affil{European Southern Observatory, Karl-Scwartzschild Stra\ss e 2,
     D-85748 Garching bei M\"unchen, Germany}
\email{epancino@eso.org, pancino@bo.astro.it}    

\author{ L. Monaco}      
\affil{Dipartimento di Astronomia - Universit\`a di Bologna, via Ranzani 1,
     40127, Bologna, ITALY}
\email{lorenzo@anubi.bo.astro.it}

\author{E. Oliva\thanks{INAF - Osservatorio Astrofisico di 
        Arcetri, Largo E. Fermi 5, 50125 Firenze, ITALY }}
\affil{INAF - Centro Galileo Galilei \& Telescopio Nazionale Galileo, 
       PO Box 565, 38700 S. Cruz de La Palma, SPAIN}
\email{oliva@tng.iac.es}


\begin{abstract}

We present  $(V,I)$ photometry of two wide ($\simeq 25 \times 25$ arcmin$^2$)
fields centered on the low surface brightness dwarf spheroidal galaxies Draco 
and Ursa Minor. New estimates of the distance to these galaxies are provided 
($(m-M)_0(UMi)=19.41 \pm 0.12$ and $(m-M)_0(Dra)=19.84 \pm 0.14$)
and a comparative study of their evolved stellar population is presented. 
We detect for the first time the RGB-bump in the Luminosity Function of UMi
($V_{RGB}^{Bump}=19.40\pm 0.06$) while the feature is not detected in Draco.
Photometric metallicity distributions are obtained for the two galaxies and an
accurate analysis to determine the intrinsic metallicity spread is performed by
means of artificial stars experiments. The adopted method is insensitive to
stars more metal poor than $[Fe/H]\sim -2.5$ and it rests on the assumpion that 
the age spread in the considered populations is small (i.e. the impact of
the actual age spread on the colors of the RGB stars is negligible).
We find that, while the average
metallicity of the two galaxies is similar ($<[Fe/H]>_{UMi} = -1.8$ and 
$<[Fe/H]>_{Dra} = -1.7$) the metallicity distributions are significantly
different, having different peak values 
($[Fe/H]_{UMi}^{mod} = -1.9$ and $[Fe/H]_{Dra}^{mod} = -1.6$) and 
different maximum metallicities. We suggest that such differences may be partly
responsible for the difference in HB morphology between the two galaxies.
The intrinsic metallicity $1-\sigma$ spread is $\sigma_i=0.10$ in UMi and
$\sigma_i=0.13$ in Draco. We demonstrate that the inner region of UMi is
significantly structured, at odds with what expected for a system in
dynamical equilibrium. In particular we show that the main density peak of UMi
is off-centered with respect to the center of symmetry of the whole galaxy and
it shows a much lower ellipticity with respect
to the rest of the galaxy. Moreover, UMi stars are shown to be clustered
according to two different characteristic clustering scales, 
as opposite to Draco, which instead has a very
symmetric and smooth density profile. 
The possible consequences of this striking structural difference on our ideas 
about galaxy formation are briefly discussed.
Combining our distance modulus with the more recent estimates of the total
luminosity of UMi, we find that the mass to light (M/L) ratio 
of this galaxy may be as low as $M/L\sim 7$, a factor 5-10 lower than current
estimates.

\end{abstract}


\keywords{(galaxies:) Local Group  - galaxies: dwarf - 
galaxies: individual (Draco, Ursa Minor) - (cosmology:) dark matter - galaxies:
structure - galaxies: fundamental parameters - galaxies: stellar content}

{}{}
\section{Introduction}

The Draco ($\alpha_{2000}= 17^h 20^m 19^s$, $\delta_{2000} = 57^{o} 54.8^{'}$) 
and Ursa Minor ($\alpha_{2000}= 15^h 09^m 11^s$, 
$\delta_{2000} = 67^{o} 12.9^{'}$)
dwarf spheroidal (dSph) galaxies are the faintest known
members of the Local Group of galaxies and they are among the lowest surface
brightness members of the group \citep{m98}.  They appear to be  dominated by 
very old (age $> 8-10$ Gyr) and metal deficient ($[Fe/H]\sim -2$) stellar 
populations \cite[see][]{m98,dradeep,apadra,grill,ken,dol7},
thus they may represent a very early and elementary stage of the evolution of 
the building blocks that may have had a primary role in the
assembly of the Milky Way galaxy \cite[see][]{bfpsex}. 
Furthermore, they are reported to have the highest Mass to Light (M/L) ratio of 
any other known galaxy 
\cite[e.g., up to $M/L=300 - 1000$ for Draco, and $M/L\sim 50-100$, for UMi, see]
[]{kle01a,taft}. Hence their
stellar content may just represent the handful of baryons trapped in a system
whose true nature is that of a huge dark halo. All these
exceptional properties have made these galaxies a classical case of study.
Nevertheless, serious observational problems (e.g., the very low
surface brightness that requires very wide field photometry to obtain
statistically significant samples of members stars) have hampered our knowledge
of these intriguing stellar systems. 

Draco and UMi are twin galaxies under many aspects: they have a similar distance
from the center of the Galaxy, similar masses and luminosities, similar metal
content and are both devoid of gas (see Table~3 for a summary of their
physical parameters). 
In this context, a comparative study
performed with strictly homogeneous observational material and with the same
data reduction / data analysis techniques may reveal interesting features.
Here we present the results of a comparative analysis, performed by obtaining
well calibrated (V,I) photometry of the evolved stars of Umi and Draco over a
field of view $\sim 25 \times 25$ arcmin$^2$, under strictly homogeneous
conditions \cite[see][for a discussion of the possible problems associated 
with the comparison of non-homogeneous photometries]{mb01}. 

The plan of the paper is as follows: in \S2 we describe the observational
material, the data reduction process and the photometric calibration; in \S3 we
present the Color Magnitude Diagrams (CMD) and the results of artificial
stars experiments; \S4 is devoted to the estimate of the distance to Draco and
Umi; in \S5 we study the properties of the Red Giant Branch of the two galaxies.
In \S6 we compare the structures of Dra and UMi
and we demonstrate that the inner region of UMi is significantly structured and
that its stars are clustered according to two different characteristic
scale-lenghts. \S7 is dedicated to the discussion of our results.

\section{Observations and Data Reduction}

\subsection{Observations}

The  data were obtained  at the  3.52 $m$  Italian telescope  TNG 
(Telescopio  Nazionale  Galileo - Roque  de  los  Muchachos, La  Palma,
Canary Islands, Spain),  using DoLoRes, a focal reducer imager/spectrograph  
equipped with a
$2048 \times 2048$ pixels thinned and back-illuminated Loral CCD array
(gain$=0.97 e^-/ADU$, read-out noise  $9.0$ ADU rms).  The pixel scale
is $0.275$  arcsec/px, thus the total  field of view of  the camera is
$9.4  \times 9.4  ~arcmin^2$. The observations were carried out during 
three nights (March 19, 20 and 21, 2001), under average seeing conditions 
($FWHM\simeq  1.0 - 1.4$ arcsec). The first and third nights of the run
were photometric. The scientific exposures taken during the second night were
calibrated indirectly by using the data acquired in the photometric nights.

Each galaxy was sampled with a square mosaic of nine partially overlapping
fields, covering a total field of fiew of $25\times 25 ~arcmin^2$. For each
field two V and two I exposures ($t_{exp}=120$ s in each filter for Dra, 
and $t_{exp}=90$ s for UMi) were secured.

\subsection{Data Analysis}

All the  raw  images were  corrected  for bias  and  flat  field, and  the
overscan  region  was  trimmed  using standard  IRAF\footnote{IRAF  is
distributed by  the National  Optical Astronomy Observatory,  which is
operated by the Association of Universities for Research in Astronomy,
Inc.,   under  cooperative   agreement  with   the   National  Science
Fundation.} procedures. Each pair of images, in each
band, has been  averaged, so that the final  analysis was performed on
averaged V and I images.

The PSF-fitting procedure was performed  independently on each V and I
average image, using a version of DoPhot \citep{doph} modified by  
P.  Montegriffo at the Bologna  Observatory to  read  images in  double
precision  format. The frames were searched for sources adopting a 5-$\sigma$
threshold, and the spatial varations of the PSF were modeled with a quadratic
polynomial.
A  final  catalogue listing  the instrumental  V,I
magnitudes  for all  the  stars in  each  field has  been obtained  by
cross-correlating the V and  I catalogues. Only the sources classified
as  stars  by  the  code  have been  retained.  The  spurious  sources
erroneously  fitted  by  DoPhot  (as  cosmic  rays,  bright  background
galaxies etc.) have been removed by hand from the catalogues.

Nine different catalogs (one for each sub-field of the mosaic) were obtained 
for each galaxy. These catalogue were reported to a homogeneous photometric
system using the large sets of stars in common among the various fields. The
magnitudes of the stars in common among adjacent fields were averaged. A
homogeneous total catalogue of instrumental magnitudes and positions was
finally obtained obtained for each mosaic field.

\subsection{Photometric calibration}

The absolute calibration has  been obtained from several repeated observations
of \citet{land} standard fields, including all the stars listed in
the extended  catalogue of calibrators provided  by \citet{stet}.
The coefficents of atmospheric extintion ($C_{ext}$) were directly obtained 
by repeated observations of the same standard field at different airmasses. 
Accurate estimates of the aperture correction were obtained with a large number
of bright and isolated stars.
The final calibration equations and $C_{ext}$ are shown in Fig.~1.

\begin{figure*}
\figurenum{1}
\centerline{\psfig{figure=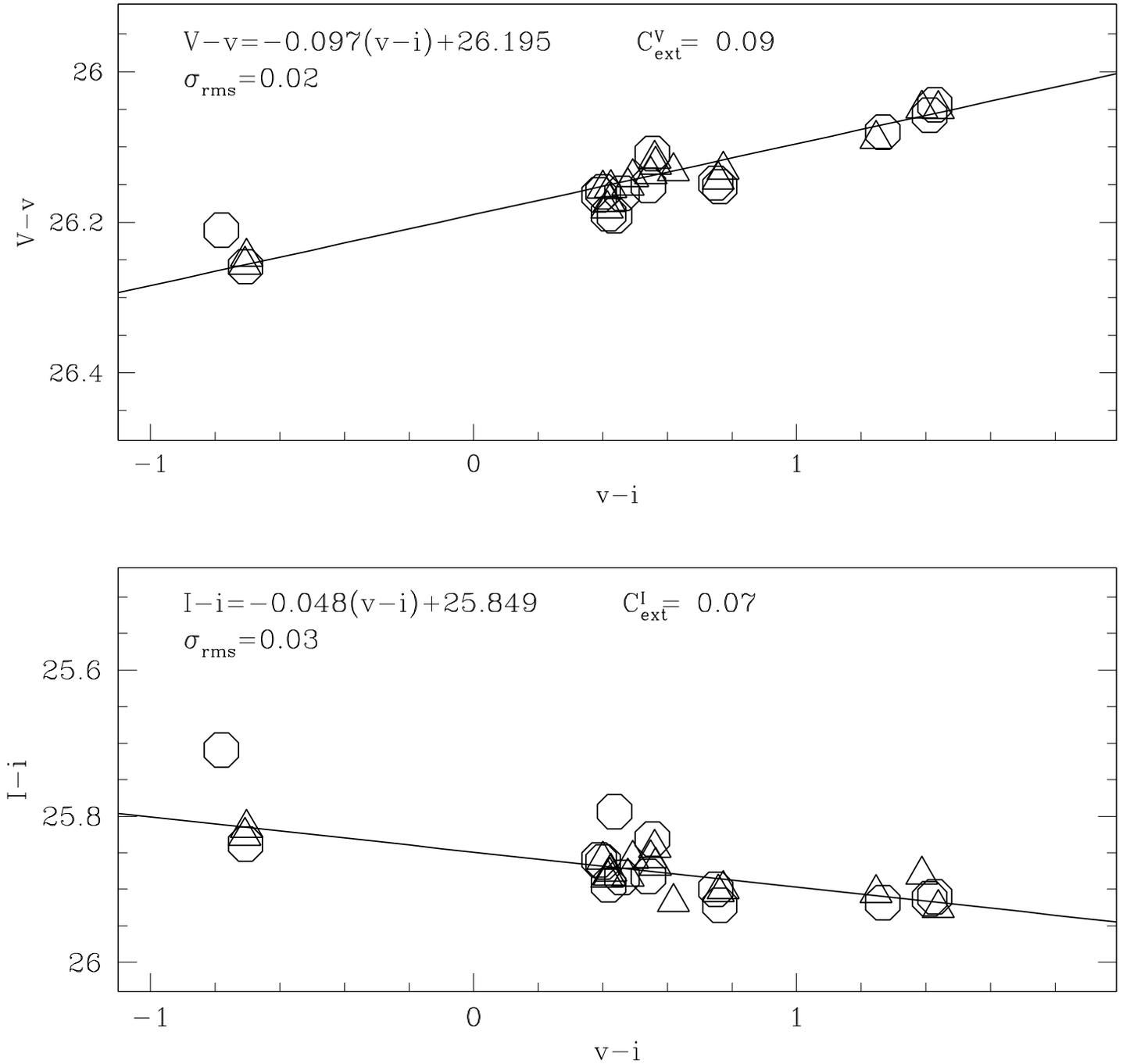}}
\caption{The difference between calibrated (upper case letters, e.g., V,I)
and instrumental (lower case letters, e.g., v,i) magnitudes for the Landolt
standard stars observed on March 19 (triangles) and March 21 (circles).
The calibration relations (upper panel: V; lower panel: I) are plotted
(solid lines) and the corresopnding equations are reported along with the root 
mean square error of the linear fits and the measuredt extinction  
coefficients.}
\end{figure*}

The accuracy of  the photometric calibration was checked by  
comparison with the independent (V,I) photometry of the Draco dSph
(kindly provided by Prof. P. Stetson, private communication), 
and with the V photometry of UMi from \cite{stet}. The results
of this test are shown in Fig~2 where the difference between
our photometry (subscript {\em t.w.}, e.g. this work) and  the one
by Stetson (subscript S00) is plotted as a function of magnitude
for both galaxies. Unfortunately no previous I band photometry of UMi
is available. The comparison   suggests that the global 
uncertainty  in the calibration is lower than $\pm 0.02$ mag in each 
passband, over the whole range of $(V-I)$  color sampled. 
The equations shown in Fig.~1 were applied to the catalogs described in \S2.2 to
produce the final calibrated catalogs that are reported in Table~1 and Table~2.
In each table we report: (column 1) the identification number, (col. 2 and 3)
the V magnitude and its error as estimated by DoPhot (see \S3.1, below), 
(col. 4 and 5) I magnitude and error, (col. 6 and 7) the position in the mosaic
in pixels, (col. 8 and 9) the position in the sky in equatorial coordinates at
the equinox 2000.0, (col. 10) the classification of the identified variable 
stars, if available and (col. 11) the name of the identified variable stars
according to the nomenclature indicated in the table note. 
The photometric errors as estimated from artificial stars experiments
will be shown and discussed in \S3.1.  

\begin{figure*}
\figurenum{2}
\centerline{\psfig{figure=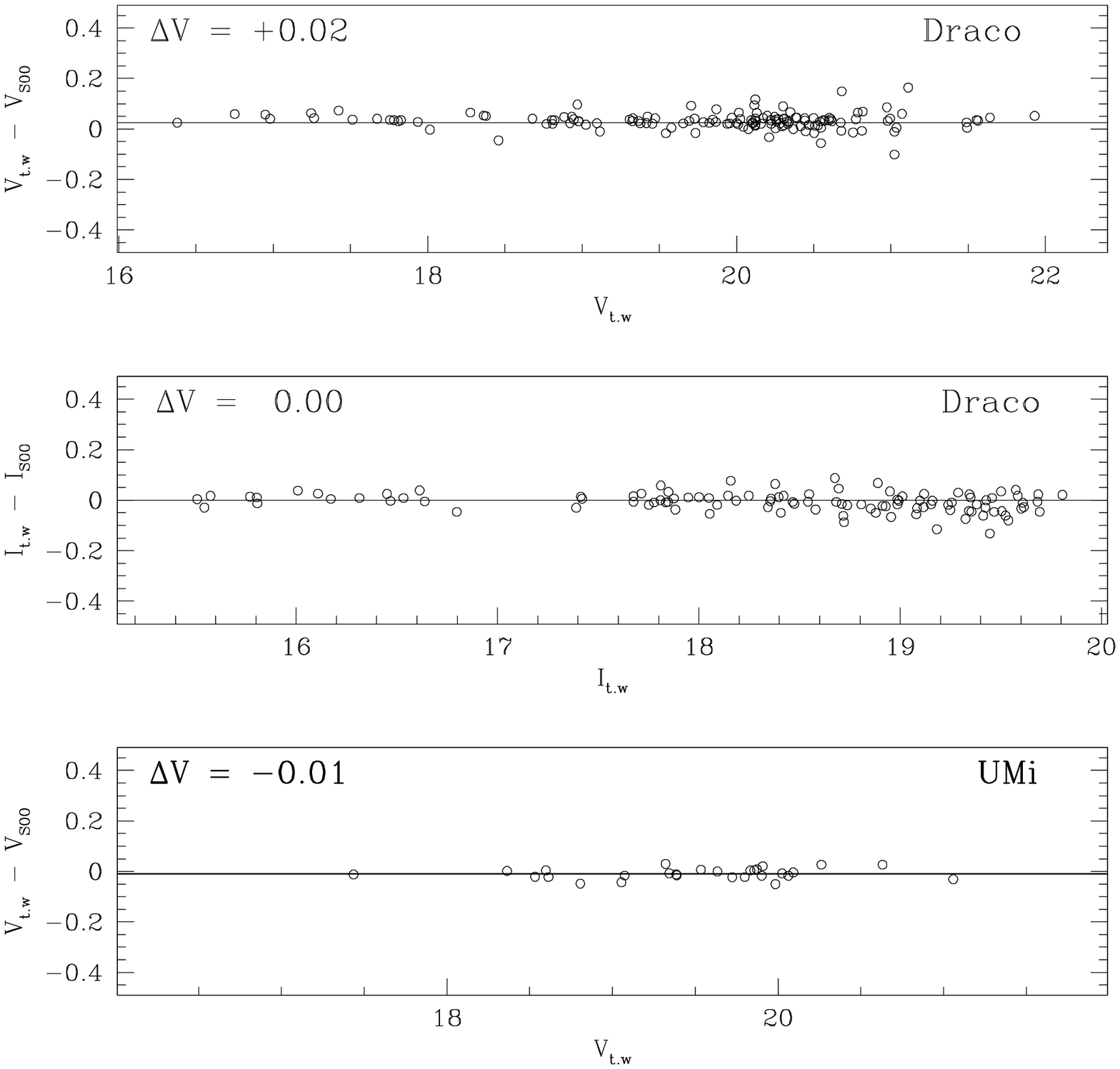}}
\caption{Difference between independently calibrated photometry from this work 
(t.w) and by \citet{stet}, for Draco (upper panel: V, middle panel: I) 
and UMi (lower panel: V).}
\end{figure*}

\section{The Color Magnitude diagram}

The final (V, V-I) Color Magnitude Diagrams (CMD) of the two galaxies are shown
in Fig.~3. In both cases our photometry reaches the base of the Red Giant
Branch (RGB) at $V\simeq 22.5$. 

The RGBs are well defined and steep, typical of
metal poor stellar systems. The RGB of UMi appears to be narrower than the
one of Draco, suggesting a smaller metallicity spread or a similar metallicity
spread but a lower mean metal content. In the CMD of UMi a hint 
of the Asymptotic Giant Branch (AGB) bump \cite[see][]{f99}
can be noted (at $V\sim 19$,$V-I\simeq
0.9$) while the corresponding feature of Draco is probably hidden by the larger
extent of contamination by foreground and background stars. Note the sharp
color cut-off of the distribution of field stars at $V-I\simeq 0.8$,
corresponding to the Main Sequence Turn Off (MSTO) of the Galactic halo/thick
disk stars \citep{heather}.
Both \citet{esk} and \citet{shet1} found that some stars lying to the
blue of the upper RGB are members of the respective galaxy, either in
Draco or in UMi. The same authors have also demonstrated that {\em all} of
these stars are not first-ascent red giants, but carbon stars instead.

Most of the Horizontal Branch (HB) stars lie in the range $19.5\le V\le 20.5$
in UMi and $20.0\le V\le 21.0$ in Draco. We have counter-identified 65
variable stars in UMi from the catalogue of \citet[][hereafter N88]{nemec}, 
that cover our whole
field of view. In Fig.~3 we have marked the type ab RR Lyrae with open squares,
the type c RR Lyrae with open triangles and the Anomalous Cepheids with open
circles, after the classification by the same authors. In Draco 56 
variable stars were
counter-identified in the central $\sim 10 \times 10~arcmin^2$ field, from the
study by \citet{baade}. Since an explicit classification is not provided by 
\citet{baade} we marked all the RR Lyrae variables with stars, while the only
Anomalous Cepheid identified is plotted as an open circle, as in the previous
case. However, it has to be recalled that, according to \citet{baade}, 
the large majority of RR Lyrae in Draco are of type ab 
\cite[see also][]{nemdra}.

\begin{figure*}
\figurenum{3}
\centerline{\psfig{figure=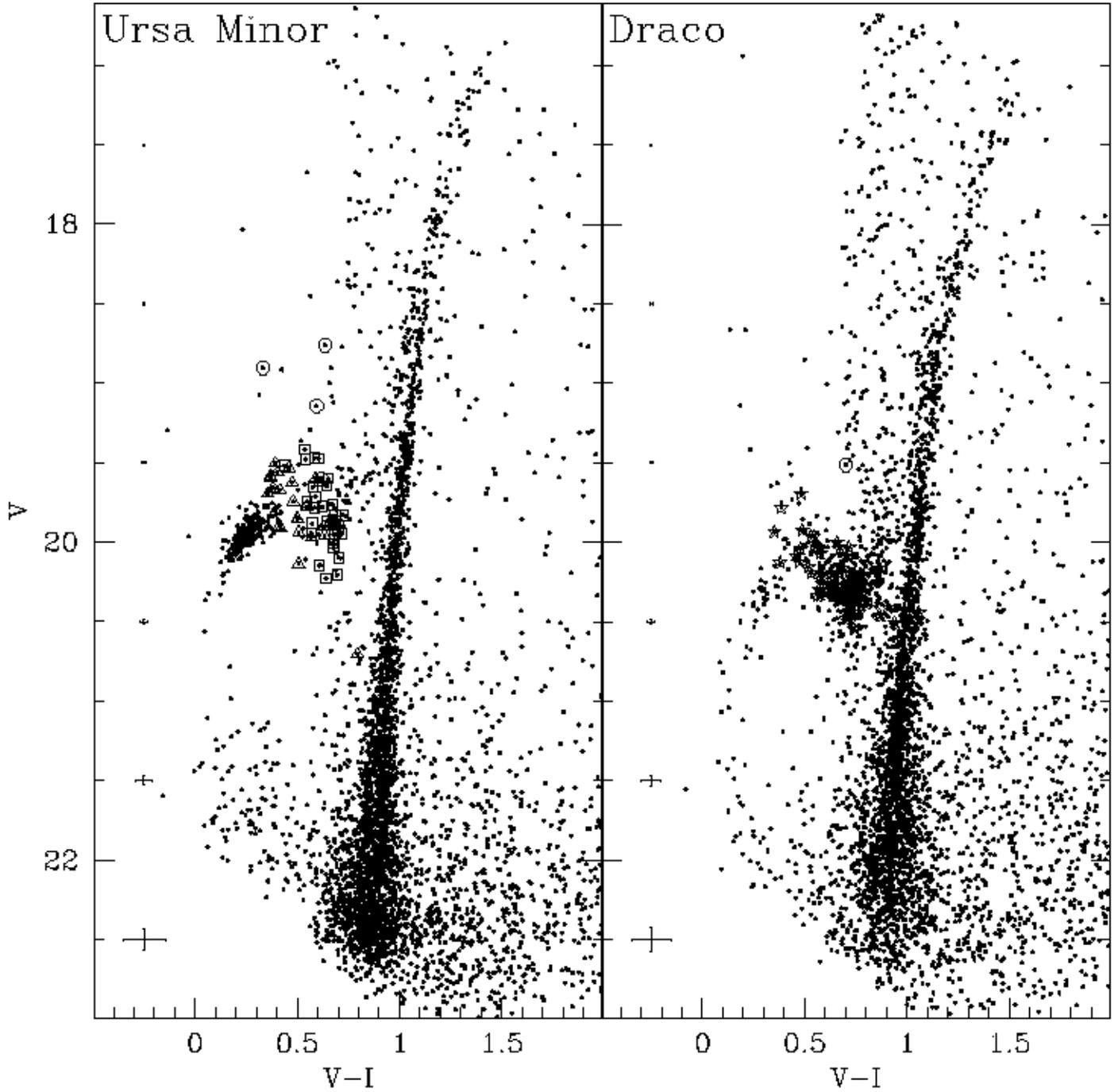}}
\caption{Color Magnitude Diagram of 4439 stars in UMi (left panel), and of 4585
stars in Draco (right panel). The Anomalous Cepheids are plotted as open
circles in both panels. In the left panel (UMi) the ab type RR Lyrae are
plotted as open squares and the c type as triangles. In the right panel
(Dra) all the RR Ly stars are plotted as stars, independently of their
type.}
\end{figure*}

As already noted by other authors the morphology of the two HBs is very 
different.
UMi has only an handful of HB stars to the red of the instability strip, a
significant number of RR Lyrae variables of both Bailey's types and a well
populated blue tail reaching $V\simeq 20.5$ and $V-I\simeq 0.0$.
On the other hand the HB of Draco is well populated in its red part and has a
sparse blue tail, reaching $V\simeq 21.0$. As can be seen from Fig.~3
there is a {\em large} number of stars in the red part of the
instability strip (see also Fig.~4, below), suggesting that the census of RR
Lyrae variables in this galaxy is far from complete. A modern CCD search for
variables over a wide field of view is urged.

Finally, at $21\le V \le 22.5$ and $0.0\le V-I\le 0.6$ the upper part of a Blue
Plume is evident in both CMDs. 
This is a well known feature of the CMD of these galaxies and
has been preferentially interpreted as a sequence of genuine Blue Straggler
stars, e.g., the result of the evolution of binary stars 
\citep{dradeep,grill,ken,car02,apadra}.

\subsection{Artificial stars experiments}

We have performed extensive artificial stars experiments in the central field of
each mosaic, i.e. the fields with the highest stellar density. We take the
results from these fields as representative of the whole mosaics.
More than 5000
stars per field have been extracted from a Luminosity Function (LF) similar to
the observed one, and with colors lying on the observed RGB ridge line 
of the galaxies, following the methods described in detail in 
\citet{mb02a,mb02b}.
The artificial stars have been simultaneously added to the V and I frames, 
$\sim 400$ at a time, with a minimum distance among them of $\sim 40$ px, 
to avoid undesired interference among artificial stars 
\cite[see][for details and references]{mb02b}.
The stars are simulated with the observed Point Spread Function and the 
complete data-reduction process has been repeated for all the frames
``enriched'' with the artificial stars.

\begin{figure*}
\figurenum{4}
\centerline{\psfig{figure=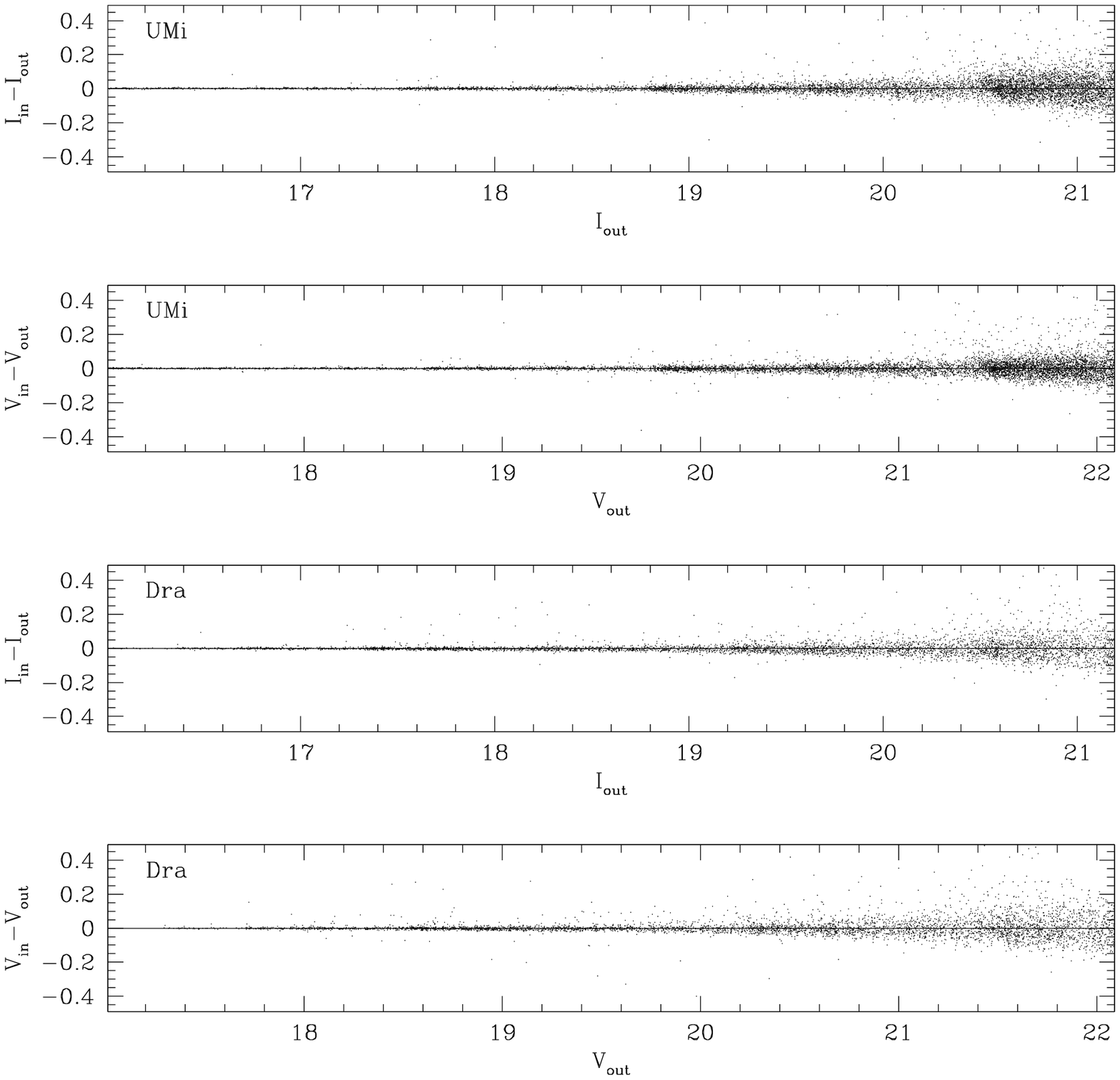}}
\caption{Magnitude residuals from artificial stars experiments}
\end{figure*}

In the present context we performed the artificial stars experiment to
obtain:

\begin{enumerate}

\item A realistic estimate of the photometric error as a function of magnitude,
      by comparing the {\em a priori} known input magnitudes with the 
      measured (output) magnitudes. This result is shown in Fig.~4.
      The average error in lower than 0.05 mag in both passbands for $V\le 20.0$
      and $\le 0.1$ for fainter magnitudes.

\item The effect of the ``observation plus data reduction'' process on the
      measured magnitudes and colors of stars lying along the RGB of a Simple 
      Stellar Population 
      \cite[SSP; e.g. a population of stars all having the same age and chemical
      composition; see][]{rf88}. This result (shown in the upper panels of
      Fig.~5) will be very useful to discriminate the color spread due to
      the intrinsic metallicity dispersion from the observational scatter.
      
\item The completeness factor ($C_f = N_{rec}/N_{inp}$, i.e. the ratio between
      the artificial stars correctely recovered and measured and the total
      number of simulated stars) as a function of magnitude.
      The lower panels of Fig.~5 show that, in both mosaics, the completeness is
      very high ($C_f>90$ \%) down to $I\sim 20.5$.      
      
\end{enumerate}

\begin{figure*}
\figurenum{5}
\centerline{\psfig{figure=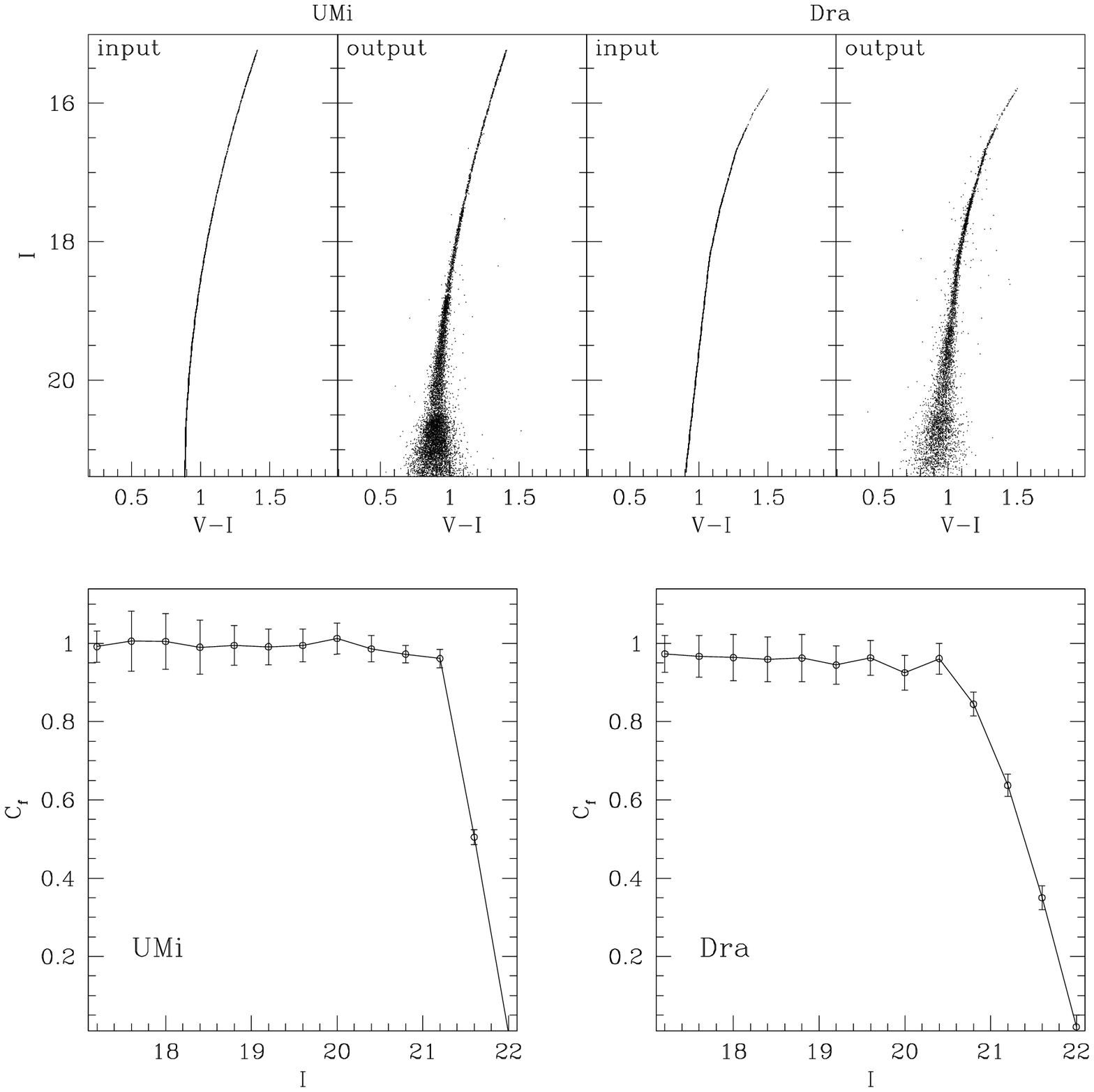}}
\caption{Upper panels: the input and output CMDs of the artificial stars for
UMi (left pair of panels) and Dra (right pair of panels). These plots show the 
effects of the {\em observation + data reduction} process on the RGB stars of a
Simple Stellar Population. Lower panels: The completeness factor ($C_f$) as a
function of I magnitude for UMi (left panel) and Dra (right panel).}
\end{figure*}

\section{The distance to UMi and Dra}

\subsection{Relative distance}

As a first step we determine the relative distance modulus of the two 
galaxies. Since they two are affected by the same
amount of interstellar extinction \cite[E(B-V)$=0.03$][]{m98,sf} and they 
share a similar metal content \citep{m98}, such
estimate can be obtained by finding the best match between the HBs.
The upper panel of Fig.~6 shows a zoomed view of the HB of UMi. The RR Lyrae
are plotted as open stars. The stars not recognized as variables by
N88 but lying in the range of color and magnitude covered by the RR
Lyrae sampled at random phase are plotted as dots, while the stars to the
red and to the blue of this region are plotted as solid circles. The solid
line in the instability strip is the mean RR Lyrae level 
($<V_{RR}>=19.86 \pm 0.07$) according to N88. The solid heavy line
is the ridge line tracking the mean of the HB distribution of UMi. The
dashed lines are displaced by $\Delta V = \pm 0.1$ mag from the ridge line and
roughly bounds all the non-variable HB stars of UMi..

\begin{figure*}
\figurenum{6}
\centerline{\psfig{figure=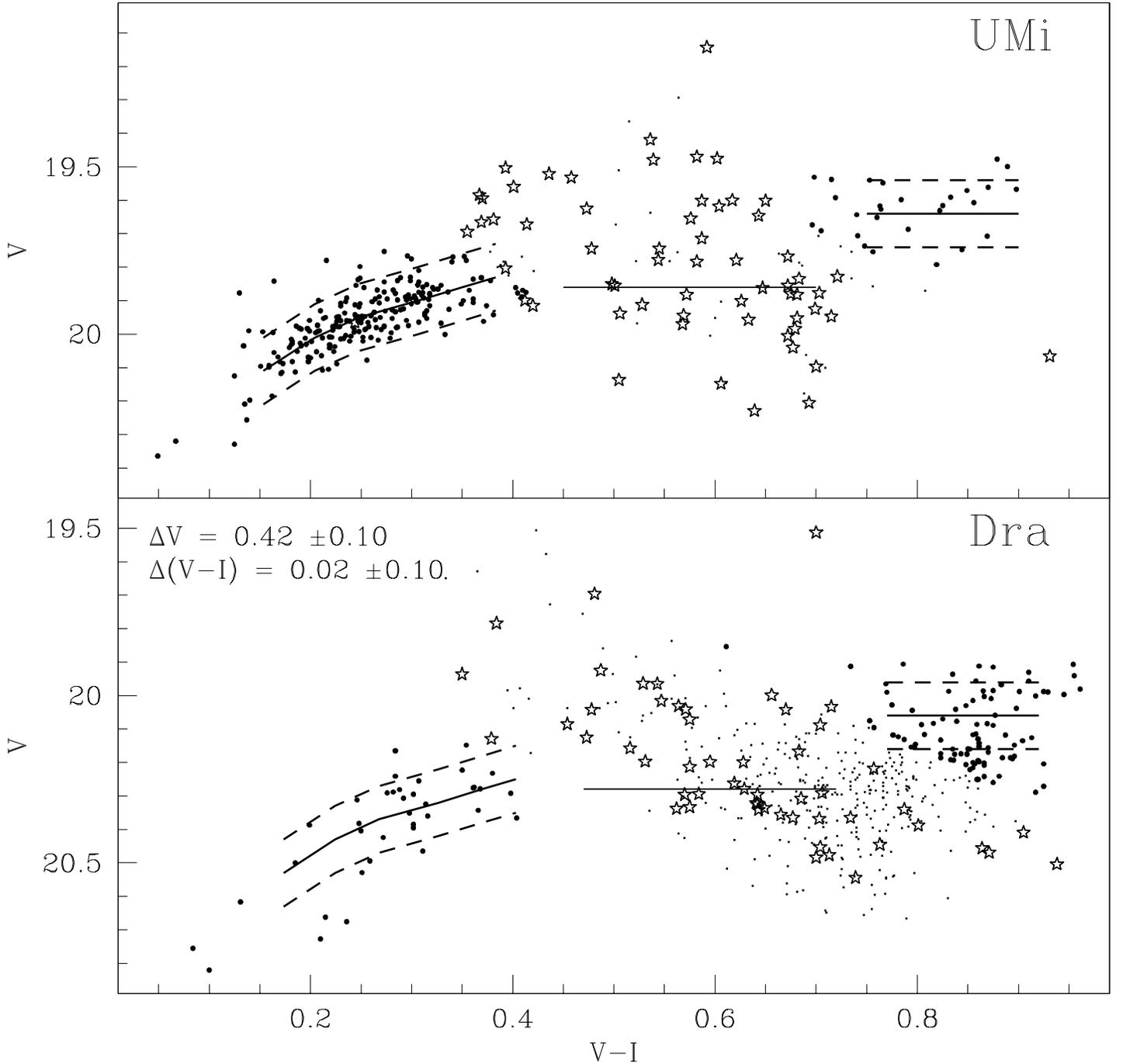}}
\caption{
Upper panel: The HB of UMi. The open stars are known RR Lyrae stars, while
the dots are stars not recognized as variables but that appear to lie in
the same region occuped by the RR Lyrae sampled at random phase. The stars to
the red and to the blue of this region (which approximates the instability
strip) are plotted as solid circles. The horizontal line in the instability
strip is the $<V_{RR}>$ level as derived by N88, and the havy line
is the fiducial for the non-variable HB stars. The dashed lines are displaced
by $\Delta V = \pm 0.1$ mag from the fiducial line. Lower panel: the HB of
Draco is plotted with the same symbols used in the upper panel. The fiducial
lines of UMi have been shifted by the reported amount to match the HB of Draco.
}
\end{figure*}

In the lower panel of Fig.~6 the HB ridge line of UMi has been
shifted to match the HB of Draco, that has been plotted with the same symbols
used above. The fit is made difficult by the different HB morphology, but the
solution obtained here with a shift of $+0.42$ mag in V and $+0.02$ in V-I  
appears as the best possible match. The derived relative distance modulus is
$\delta_{(m-M)}=0.42\pm 0.10$. The result is in good
agreement with the difference of moduli that can be obtained from Table 2
of \citet{m98}, i.e. $\delta_{(m-M)}=0.47 \pm 0.18$.

\subsection{Comparison with template clusters}

An indirect estimate of $<V_{RR}>$ and of the distance modulus can be obtained
by finding a match with the HB of well studied template clusters of similar
metallicity, for which a robust direct estimate of $<V_{RR}>$ is available
\cite[see, e.g.][and references therein]{mont98}.

In Fig.~7 we report the results of the comparison with the photometry of
M~68 ($[Fe/H]=-2.09$) by \citet{w94}. The HB of UMi (upper panel) and Draco
(lower panel) are plotted with the same symbols as Fig.~6, while the stars of 
M~68, shifted to match the HB of UMi and Dra, are represented by open circles. 
The solid line is the mean level of the
RR Lyrae in M~68 \cite[$<V_{RR}>=15.64 \pm 0.01$, according to][]{w94} after
the application of the shift needed to provide the match.

A good match is found between the HB of M~68 and UMi applying a shift of
$4.24$ mag in V to the stars of M~68. Using the above quoted $<V_{RR}>$ 
of M~68, and propagating all the
uncertainties $<V_{RR}>(UMi)=19.88 \pm 0.10$ is inferred, in excellent
agreement with the estimate by N88. We derive also the level of the
Zero Age Horizontal Branch ($V_{ZAHB}$, a quantity that is best suited for
comparison with theoretical models) by applying the same $\Delta V$ to the
$V_{ZAHB}$ value of M~68 reported by 
\citet[][hereafter F99]{f99}, $V_{ZAHB}=15.75\pm 0.05$. We obtain 
$V_{ZAHB}(UMi)=19.99 \pm 0.11$. Finally, taking into account the difference in
reddening and adopting $(m-M)_0(M68)=15.11\pm 0.1$ from F99 we obtain the
following estimate for the distance modulus of UMi: 
$(m-M)_0=19.38 \pm 0.16$, inclusive of all the uncertainties.

\begin{figure*}
\figurenum{7}
\centerline{\psfig{figure=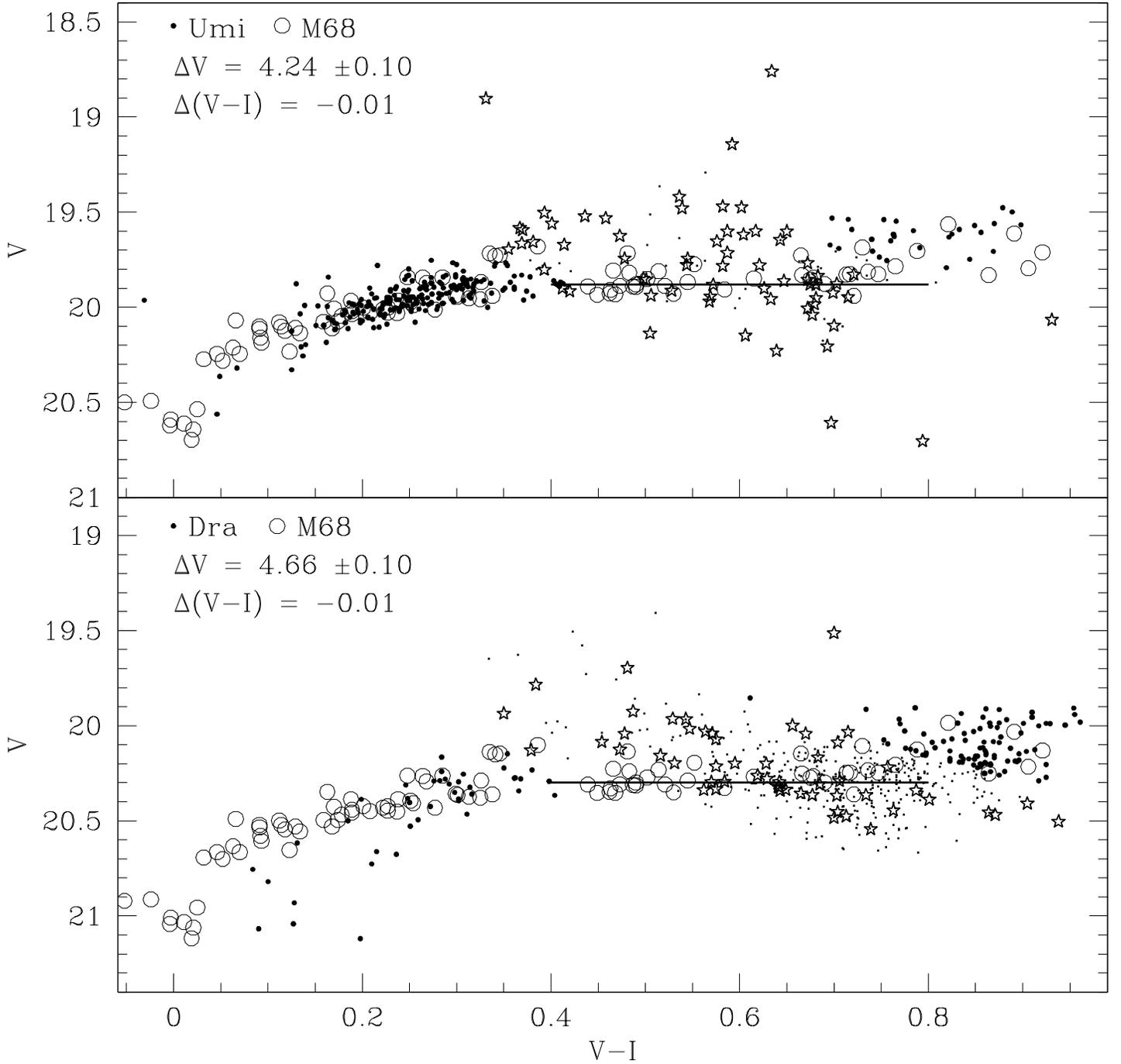}}
\caption{Matching the HBs of UMi (upper panel) and Draco (lower panel) with the
HB of M~68. The symbols for UMi and Draco are the same as in Fig.~6 while the
M~68 stars are plotted as open circles. The solid line is the average RR Ly
level of M~68 sfithed by the amount specified in the upper left corners of the
panels.}
\end{figure*}

The HB morphology of Draco has no counterpart among the galactic globular 
clusters of similar metallicity. Hence, the fit with the
HB of M~68 - shown in the lower panel of Fig.~7 -is worse than that found for
UMi and do not provide a well constrained solution. 
However, with the adopted shift we obtain results that are 
in full agreement with the differential distance between UMi and Draco 
derived in \S4.1. With the same procedure described above we obtain for Draco:
$<V_{RR}>=20.30 \pm 0.12$, $V_{ZAHB}=20.41\pm 0.13$ and
$(m-M)_0=19.80 \pm 0.18$

In Fig.~8 are reported the results of the comparison with the HST photometry of
M~92 ($[Fe/H]=-2.24$) by Ferraro and co-workers (Ferraro et al., 
in preparation), that is also in good agreement with 
the well calibrated ground-based photometry by \citet{boltem92}. 
The symbols and the procedure are the same adopted in Fig.~7
for the comparison with M~68.

Also in this case a good match is obtained for UMi (upper panel), with the 
shifts reported in the figure. We adopt larger uncertainties in the derived
quantities, with respect to the case of M~68, since M~92 is slightly more metal
poor than UMi (and Draco). Adopting 
$<V_{RR}>(M~92)=15.08 \pm 0.01$ from \citet{kop} we obtain 
$<V_{RR}>(UMi)=19.84 \pm 0.12$, again in excellent agreement with N88
as well as with the results from Fig.~7. Adopting 
$V_{ZAHB}(M~92)=15.30\pm 10$ from F99 it is obtained 
$V_{ZAHB}(UMi)=20.06\pm 0.14$, in agreement with what found above. Finally,
adopting $(m-M)_0(M~92)=14.74 \pm 15$ from F99, it is obtained 
$(m-M)_0(UMi)=19.47 \pm 0.20$.

\begin{figure*}
\figurenum{8}
\centerline{\psfig{figure=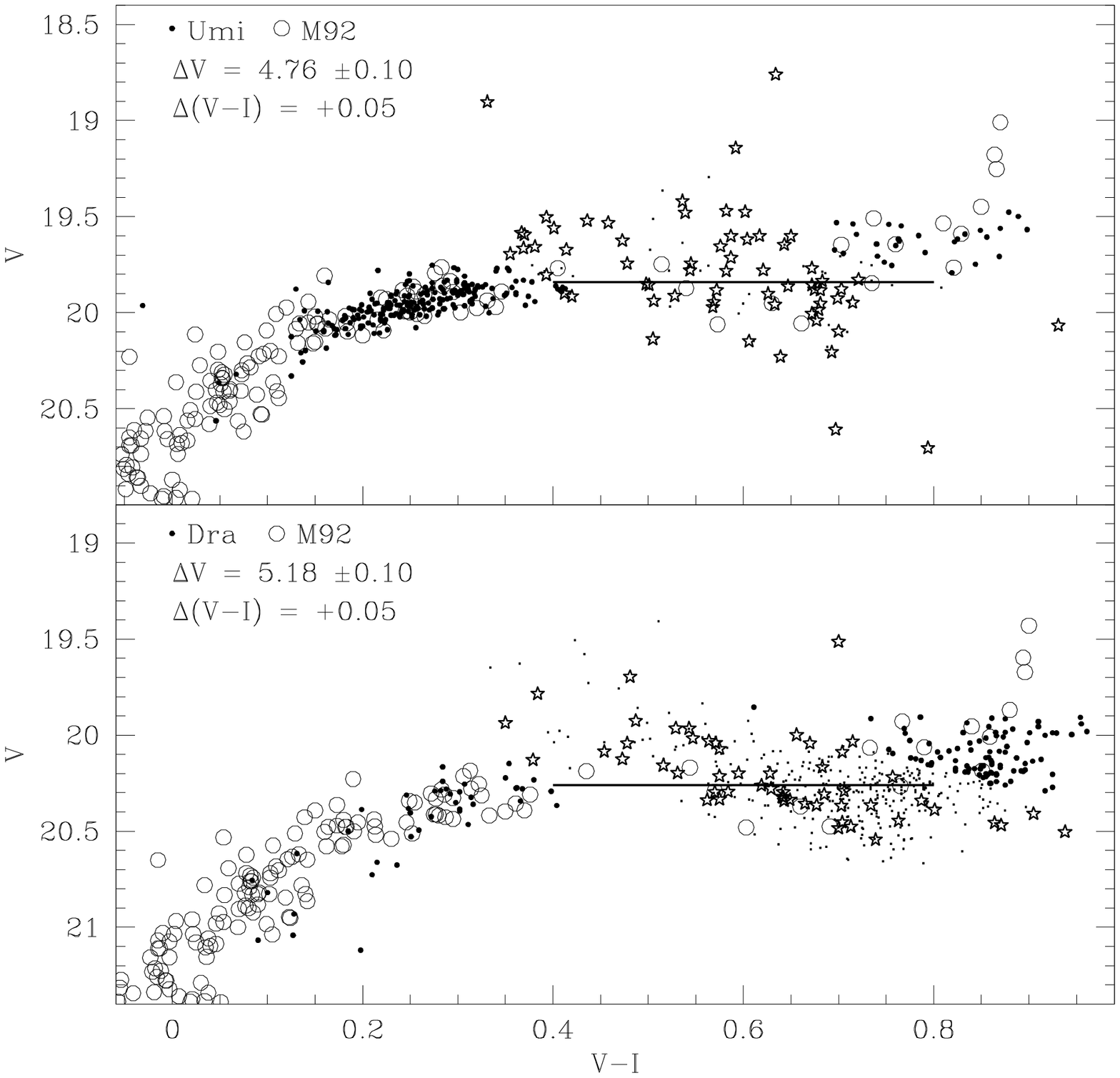}}
\caption{Matching the HBs of UMi (upper panel) and Draco (lower panel) with the
HB of M~92. The symbols for UMi and Draco are the same as in Fig.~6 and 7
while the M~92 stars are plotted as open circles. The solid line is the average 
RR Ly level of M~92 sfithed by the amount specified in the upper left 
corners of the panels.
}
\end{figure*}

M~92 appears to provide a better match to the HB of Draco, with respect to M~68
(Fig.~8, lower panel). With the same assuptions as above, we obtain for Draco:
$<V_{RR}>= 20.26 \pm 0.12$, $V_{ZAHB}=20.48\pm 0.14$ and 
$(m-M)_0(Dra)=19.89 \pm 0.20$. The overall agreement with the results obtained
from the comparison with M~68 is fully satisfying.

In conclusion, we adopt the weighted mean of the results obtained above for
$<V_{RR}>$, $V_{ZAHB}$ and $(m-M)_0$. The final results are 
$<V_{RR}>=19.86 \pm 0.09$, $V_{ZAHB}=20.02 \pm 0.09$, for
UMi, and $<V_{RR}>=20.28 \pm 0.10$, $V_{ZAHB}=20.44 \pm 0.10$, for Draco.
The adopted distance moduli are $(m-M)_0(UMi)=19.41 \pm 0.12$ and 
$(m-M)_0(Dra)=19.84 \pm 0.14$

\subsection{A consistency check: the RGB Tip}

The use of the Tip of the Red Giant Branch (TRGB) as a standard candle is a
powerful technique to estimate the distances to galaxies that host old
stellar populations \citep[see][and refereces therein]{tiprev}. \citet{bfptip}
have recently provided a new robust calibration of the zero-point of the
relation between the absolute I magnitude of the tip ($M_I^{TRGB}$) and the
metallicity that is independent of the distance scale based upon classical
standard candles (RR Lyrae, Cepheids). This new calibration prompted us to start
a large observational programme aimed at a re-assessment of the distances
to Local Group galaxies based on the TRGB technique.
The present study is the first of a series describing the
results of this program. Unfortunately, despite the large field sampled, UMi and
Draco have too few stars (and a too low surface brightness) to allow an
individual safe application of the TRGB technique 
\cite[see][hereafter MF95]{smf96,tipsim}.
However, it is of great interest for the final scope of the project to
check if the distance of these galaxies is consistent with the new TRGB
distance scale.

\begin{figure*}
\figurenum{9}
\centerline{\psfig{figure=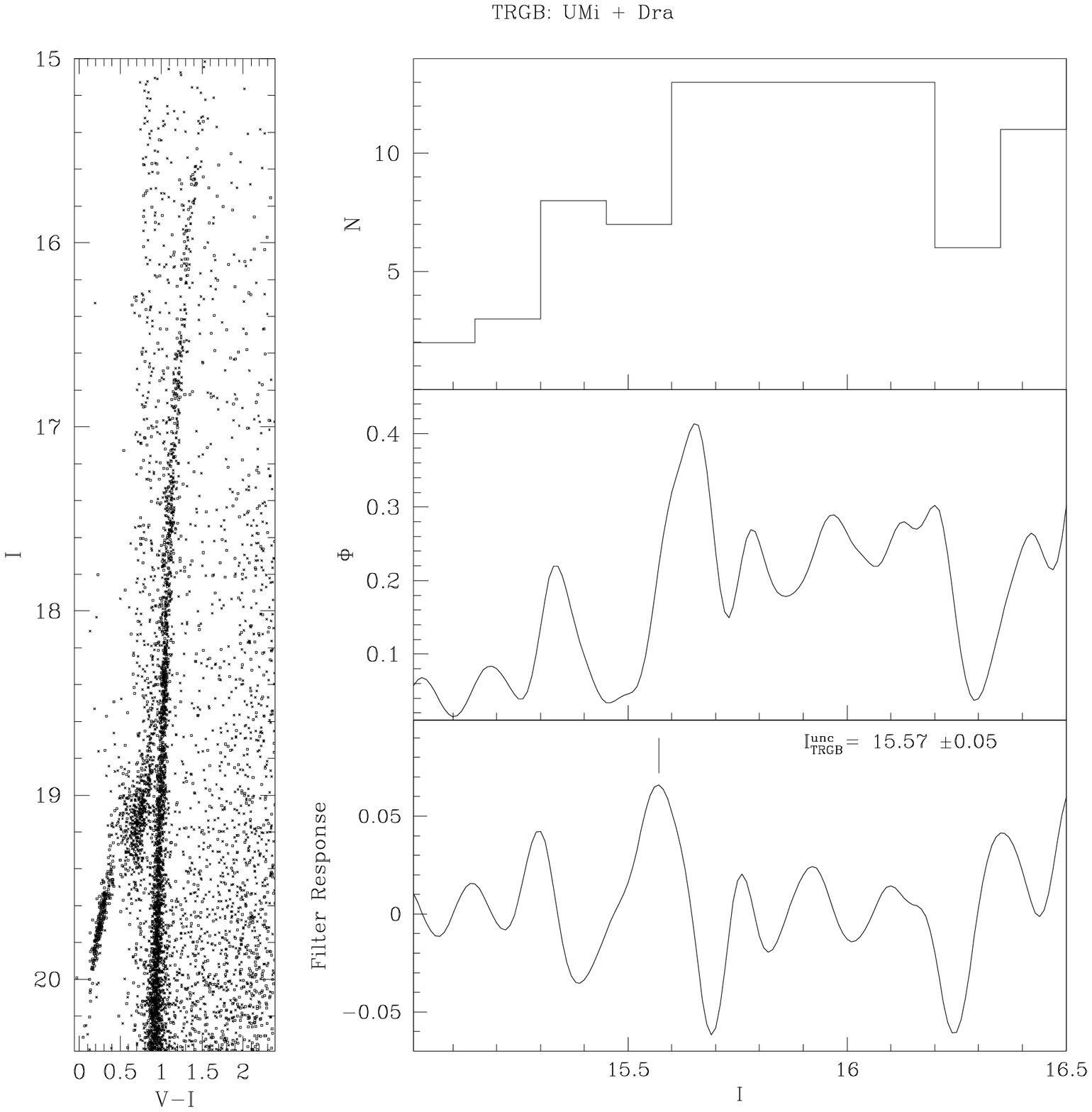}}
\caption{Left panel: composite CMD of UMi and Draco, obtained by shifting the
magnitudes of Draco stars by -0.42 mag. Right panels: ordinary histogram (upper
panel), generalized histogram (middle panel) and Sobel's filter response (lower
panel) for the LF of the upper RGB of the composite sample. The position of the
detected TRGB is indicated in the lower panel. 
}
\end{figure*}

To do this check we adopt the shift derived in \S4.1 to move Draco at the
distance of UMi and we try to detect the TRGB from the merged sample, having now
a number of stars nearly double with respect to each individual
case. In the left panel of Fig.~9 we report the composite (I;V-I) CMD of the two 
galaxies together, after the application of the quoted shift to Draco. 
The right panels
of Fig.~9 report the actual detection of the TRGB using the standard technique,
as defined by \citet{smf96}. The sharp cut-off of the Luminosity Function (LF)
of the RGB (that is the actual observational marker of the TRGB) is clearly
visible in the histograms shown in the upper and middle right
panels. The edge-detector Sobel filter shows the TRGB as the most pronounced
peak at I=$15.57\pm 0.05$. The considered sample has 100 stars in the
upper 1-magnitude bin, however, from the CMD of field stars we estimate 
that $\simeq 20$ \% of them are likely 
foreground sources. Hence the actual number of stars in the uppermost 1-mag bin is in fact
$N_{\star}\simeq 80$. By means of numerical simulations MF95 showed
that a reasonably safe detection of the TRGB can be made only if
$N_{\star}>100$, thus it is likely that our detection overestimates the actual I
magnitude of the tip. To quantify this systematic error we repeat the MF95
numerical experiment. 
We adopted the ridge line of M~68 as an RGB template and a
model RGB-LF of the same form as the one used by MF95, imposing a cut-off at
$M_I=-4.00$. Then we randomly generated 100 synthetic RGBs for each of the
following cases: $N_{\star}=20,40,60,80,100,150,200$, and we detected the tip of
each synthetic RGB with the standard technique. Observational errors extracted
from a gaussian distribution with $\sigma_V,\sigma_I$ similar to those of the 
observed
stars were added to each synthetic star before the TRGB detection. The average
$M_I(TRGB)$ of the 100 simulated RGBs and the corresponding standard deviations
are plotted as a function of $N_{\star}$ in Fig.~10. The results by MF95 are
confirmed: for $N_{\star}\ge 100$ the estimate from the standard technique is
within $\le 0.02$ mag from the {\em true} TRGB and the standard deviation is
$\le 0.04$ mag. The efficiency of the method drops suddenly for 
$N_{\star}\le 80$ and the associated uncertainty becomes $\simeq 0.1$ or larger.
The difference between the {true} luminosity of the tip and the observed one can
be modeled with a first order polinomial as a function of $N_{\star}$, for 
$N_{\star}< 100$:

$$\delta I_{TRGB} = I_{TRGB}^{true} - I_{TRGB}^{obs} = 0.0022 N_{\star} 
- 0.236$$

In the present case, with $N{\star}\simeq 80$, 
$\delta I_{TRGB} = -0.07\pm 0.08$, thus
our final estimate for the TRGB location is $I_{TRGB}= 15.50 \pm 0.14$, where we
have considered also the uncertainties associated with the correction and with
the Dra-UMi shift we have applied to  obtain the considered merged sample.
The galaxies are dominated by a metal poor population quite similar to that of
the template used by \citet{bfptip}, that obtained $M_I^{TRGB}= -4.04\pm 0.12$
at $[Fe/H] = -1.7$, for the globular cluster $\omega$ Centauri. 
Adopting $E(B-V)=0.03$, we obtain  
$(m-M)_0 = 19.50 \pm 0.20$ and $(m-M)_0 = 19.92 \pm 0.27$ for UMi and 
Draco respectively, in good agreement with the results of \S4.2. 

\begin{figure*}
\figurenum{10}
\centerline{\psfig{figure=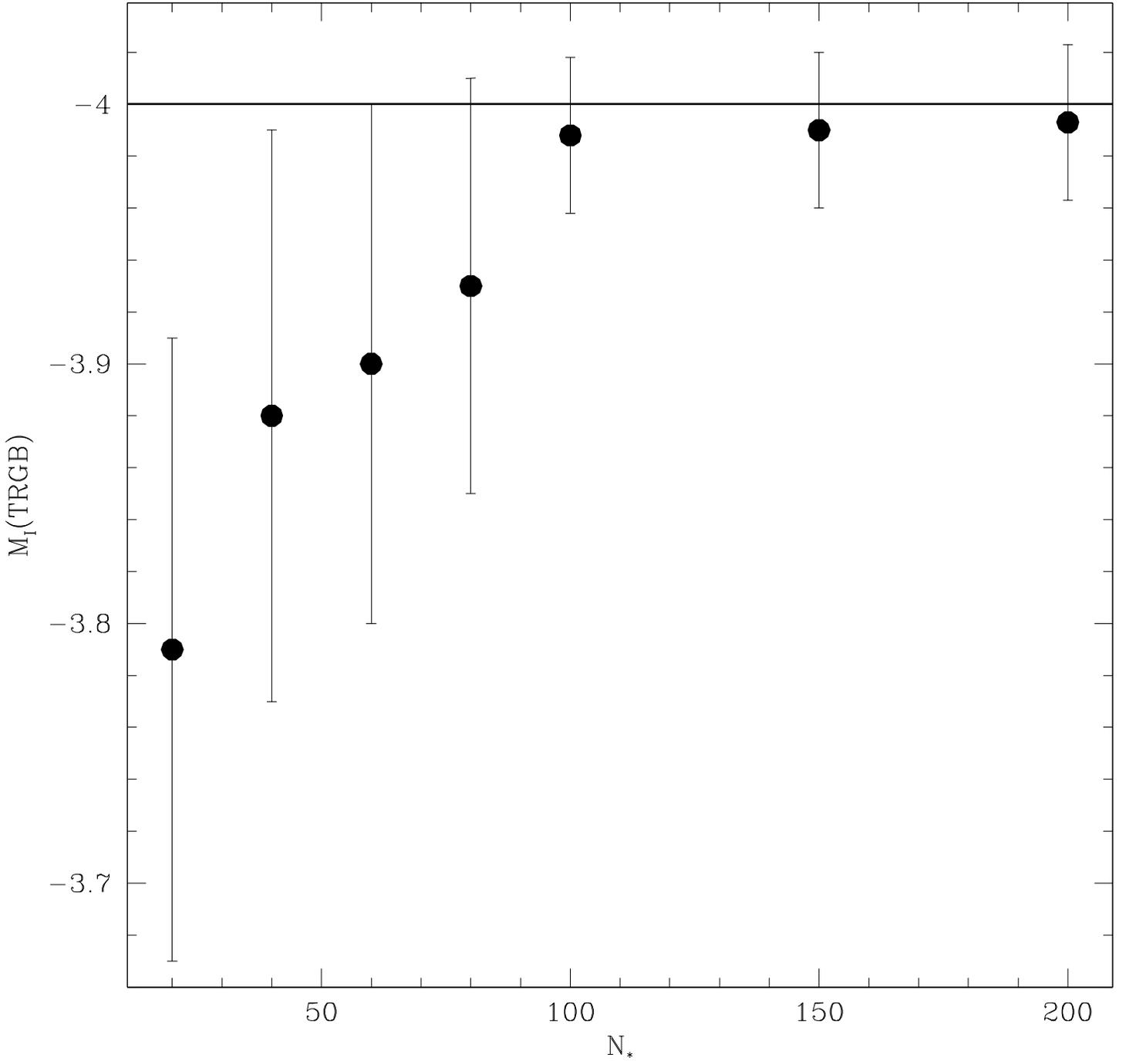}}
\caption{Detected absolute I magnitude of the TRGB as a function of the number
of stars in the upper 1 mag bin. Each point is the average of 100 simulated
samples having $N_{\star}$ stars in the upper bin. The error bars are the
corresponding standard deviations.
}
\end{figure*}

Hence we conclude that the distance moduli derived in \S4.2 are fully consistent
with the TRGB distance scale as calibrated by \citet{bfptip}. 

\section{The stellar content of Umi and Draco}

\subsection{The RGB bump}

The RGB-Bump is an evolutionary  feature occurring along the RGB which
flags  the  point  where  the  H-burning shell  crosses  the  chemical
discontinuity  left  by  the  maximum penetration  of  the  convective
envelope. Theoretical models predict that the position of the RGB-Bump is
mainly driven by metallicity, with a mild dependence on age (see F99 and
references therein). 
 
From the  observational point of view the feature was observed for the first
time in a globular cluster by \citet{kbump} and it  was only recently
identified in     a     significant     number     of     clusters
\citep[][F99]{ffp90,manu99}. The first
detection of the RGB-Bump in a galaxy is very recent too. 
\citet{sculp} identified
a double RGB-bump in the Sculptor dwarf spheroidal and took this evidence as
indicative of the presence of two populations of different metallicities. A very
similar feature has been revealed also in the Sextans dSph by
\citet{bfpsex}. At present, these are the only two detections of the RGB-bump 
ever obtained in stellar systems other than globulars.

The change in the slope of the cumulative luminosity function (LF) and
the excess  of star counts in the  differential LF of the  RGB are the
main  tools  to identify  the  RGB-Bump.  In particular  \citet{ffp90}
suggest  that the  change in  the slope  of the  cumulative LF  is the
safest indicator of the RGB-Bump. In Fig.~ 11 the differential and cumulative 
LFs of the RGB in the region of the bump are shown for both galaxies.

\begin{figure*}
\figurenum{11}
\centerline{\psfig{figure=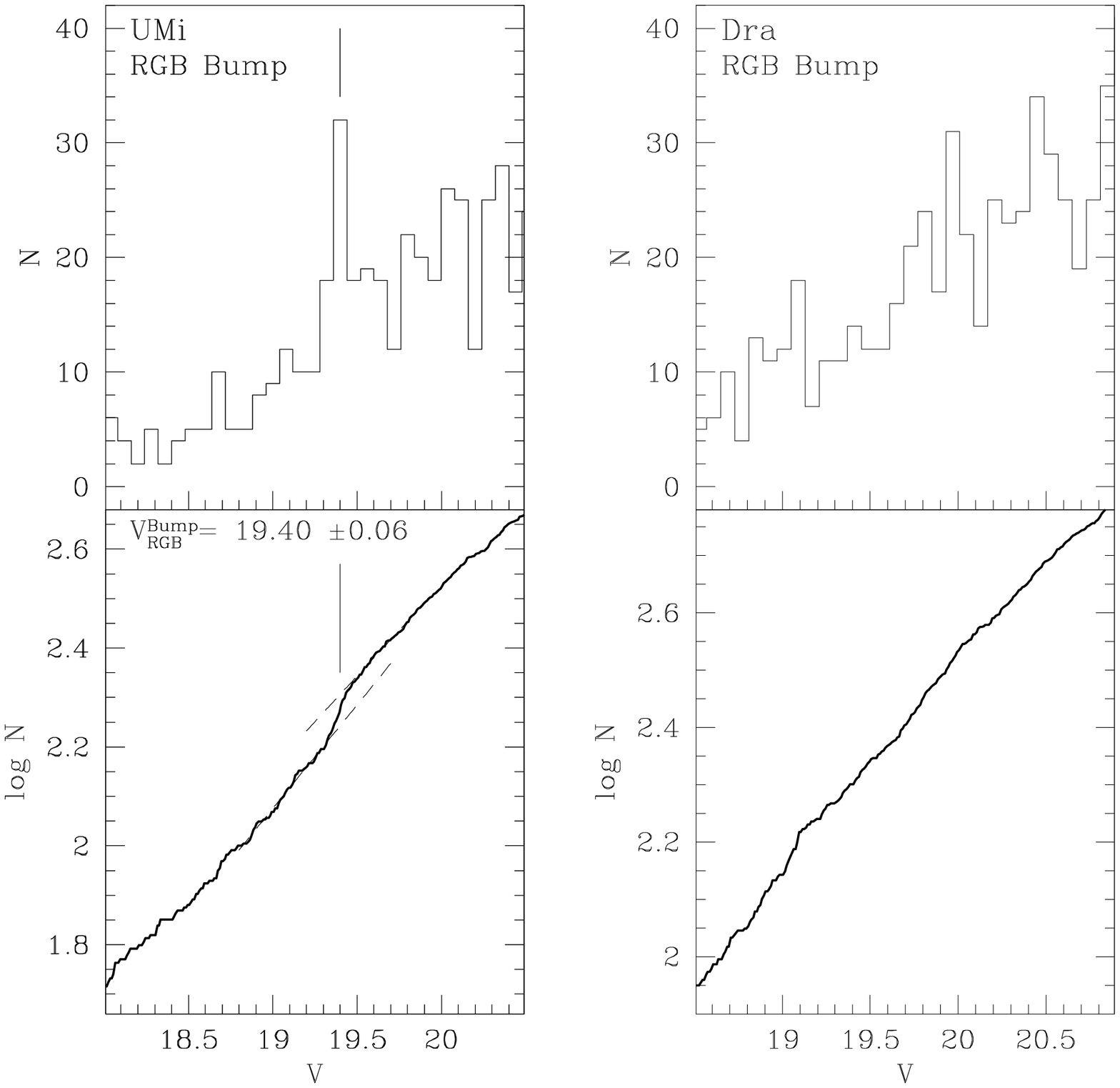}}
\caption{Differential (upper panels) and cumulative (lower panels) LF of the RGB
of UMi (left panels) and Draco (right panels). The position of the RGB-bump of
UMi is clearly indicated.
}
\end{figure*}

The RGB-bump is clearly detected in UMi at $V^{bump} = 19.40 \pm 0.06$. This is
the first {\em single} RGB-bump ever detected in a galaxy.
On the other hand, there is no sign of this feature in the LF of Draco. 
We argue that a larger spread of metallity and/or
age in the stellar population of Draco, with respect to UMi, is responsible
for the smearing of the RGB-bump feature in this galaxy (see \S5.2). 
This suggests that these dSphs - that are
quite similar under many aspects \cite[see][and references therein]{bfpsex} -
had different evolutionary histories at early times.

The magnitude difference between the RGB-bump and the  HB level ($\Delta
V_{HB}^{bump}$) has been  used by \citet{ffp90}, and later  by F99 and
\citet{manu99},  to  compare  the  observations with  the  theoretical
predictions.   The  $\Delta   V_{HB}^{bump}$  parameter  is  mainly  a
function of the  metallicity and it has only a  mild dependence on age.  
Adopting the  ZAHB levels obtained in 
\S4.2, we obtain for UMi $\Delta V_{HB}^{bump}=-0.62 \pm 0.11$. Since it has
been shown that UMi is dominated by a very old population that formed in a
rather short timescale \citep{car02,ken,umilf,dol7} we can safely use the 
$\Delta V_{HB}^{bump}$ parameter as an indicator of the mean metal content of
the old stars. We use the data by F99 to obtain a fit of $[Fe/H]$ as a
function of $\Delta V_{HB}^{bump}$, adopting the metallicity scales 
by \citet{ZW84} (ZW) and \citet{cg97} (CG). It is found:

$$ [Fe/H]_{ZW}=1.426\Delta V_{HB}^{bump} - 1.233 ~~~~~(rms=0.08)$$

$$ [Fe/H]_{CG}=-0.430\Delta {V_{HB}^{bump}}^2 + 1.183\Delta V_{HB}^{bump}  
- 1.046 ~~~~~(rms=0.07).$$

With these relations we obtain for UMi $<[Fe/H]_{ZW}> = -2.1 \pm 0.2$ and 
$<[Fe/H]_{CG}> = -1.95 \pm 0.2$, in good agreement with the estimates
obtained with other methods \cite[see][]{m98,car02,shet2}.

\subsection{The metallicity distributions}

In Fig.~12 the (I,V-I) distribution of the upper RGB of UMi (left panel) and
Draco (right panel) are compared with the ridge lines of the template globular
clusters NGC~6341, NGC~6205 and NGC~288, from left to right, taken from the
homogeneous set by \citet{ivo}. In the \citet{ZW84} metallicity scale
the templates have $[Fe/H]=-2.24, -1.65$ and $-1.40$
respectively, while in the \citet{cg97} scale their metallicities are 
$[Fe/H]=-2.16, -1.39$ and $-1.07$ respectively. The metallicities, distance
moduli and reddenings of the templates are taken from F99. 
For the two galaxies we adopted the estimates of \S4. The meaning of the
horizontal lines and of the large empty symbol will be discussed in \S5.2.1.

\begin{figure*}
\figurenum{12}
\centerline{\psfig{figure=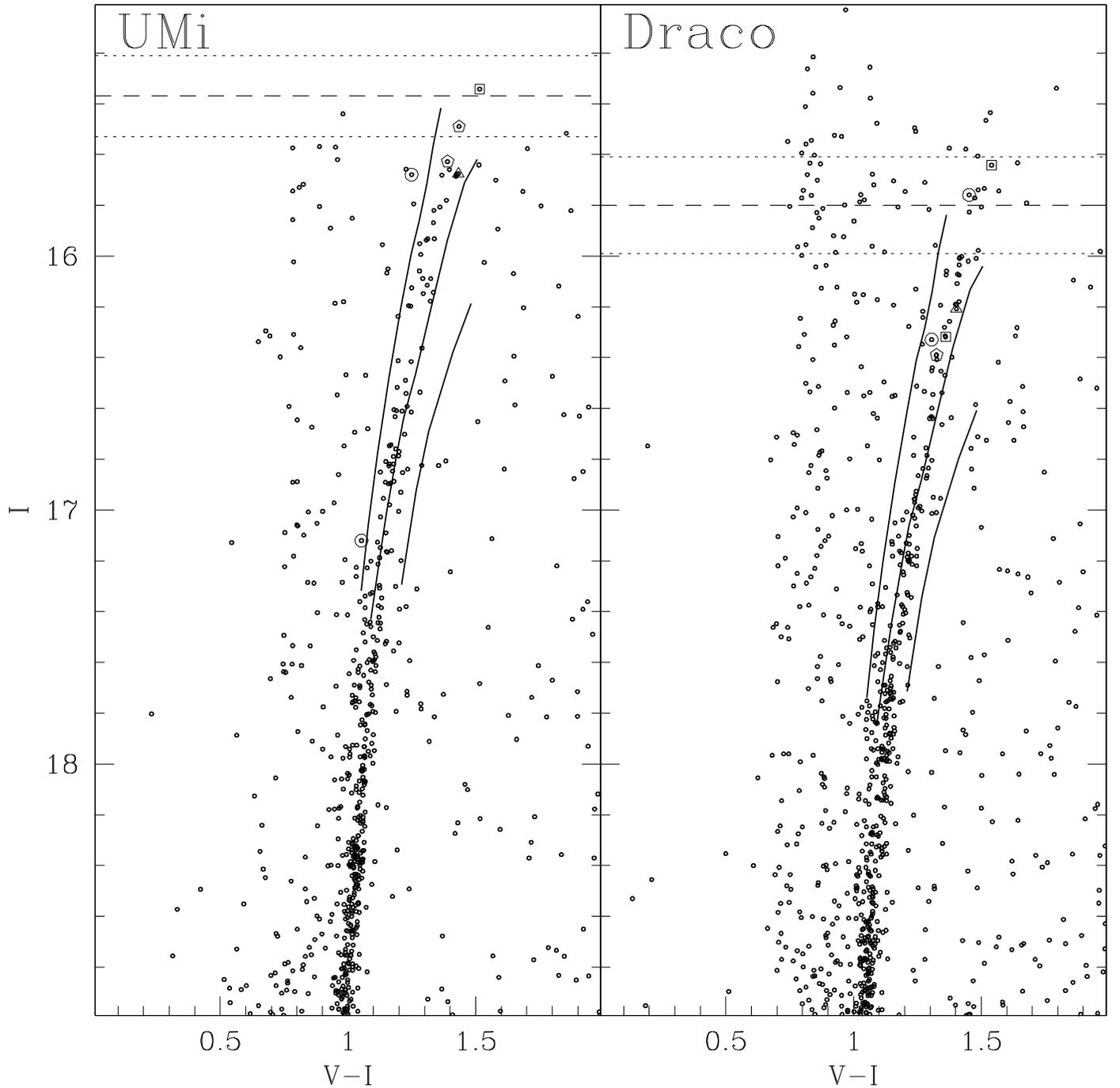}}
\caption{The (I,V-I) distribution of the upper RGB of UMi (left panel) and
Draco (right panel) are compared with the ridge lines of the templates globular
clusters NGC~6341 ($[Fe/H]_{ZW}=-2.24$; $[Fe/H]_{CG}=-2.16$), 
NGC~6205 ($[Fe/H]_{ZW}=-1.65$; $[Fe/H]_{CG}=-1.39$) and 
NGC~288 ($[Fe/H]_{ZW}=-1.40$; $[Fe/H]_{CG}=-1.07$), from left to right. 
The long-dashed line is the expected level of the TRGB with the adopted 
distance moduli (from \S4.2) and the TRGB calibration by \citet{bfptip}. 
The dotted lines enclose the full range of uncertainty in the expected 
magnitude of the TRGB. The large empty symbols marks the stars observed by
\citet{shet2} acoording to the metallicity estimated by these authors.
Circles: $[Fe/H]\le -2.10$, pentagons: $-2.10 <[Fe/H]\le -1.8$, 
squares: $-1.8 <[Fe/H]\le -1.5$, triangles: $-1.5 <[Fe/H]\le -1.3$. 
}
\end{figure*}

The large majority of the UMi RGB stars are enclosed between the ridge lines
of NGC~6341 and NGC~6205, implying a mean metallicity around 
$[Fe/H]_{ZW}\simeq -2.0$
in good agreement with the result found in \S5.1 as well as with previous
estimates, either photometric or spectroscopic 
\cite[see][and references therein]{car02,shet2}. The RGB of
Draco seems slightly redder than the one of UMi, suggesting a mean metallicity
$[Fe/H]_{ZW}\simeq -1.7$, also in agreement with previous estimates
\cite[see][and references therein]{apadra,shet2}.
Both galaxies result significantly more metal
rich (by $\sim 0.3$ dex) if the CG metallicity scale is adopted, as a 
natural consequence of the higher metallicity value of all the adopted 
templates in this scale. Independently of the assumed scale, it is clear that
the color spread along the RGB of both galaxies is significantly larger than the
photometric scatter (see Fig.~4), thus we confirm the presence of an intrinsic
metallicity spread \citep{shet2,car02,apadra}.

\begin{figure*}
\figurenum{13}
\centerline{\psfig{figure=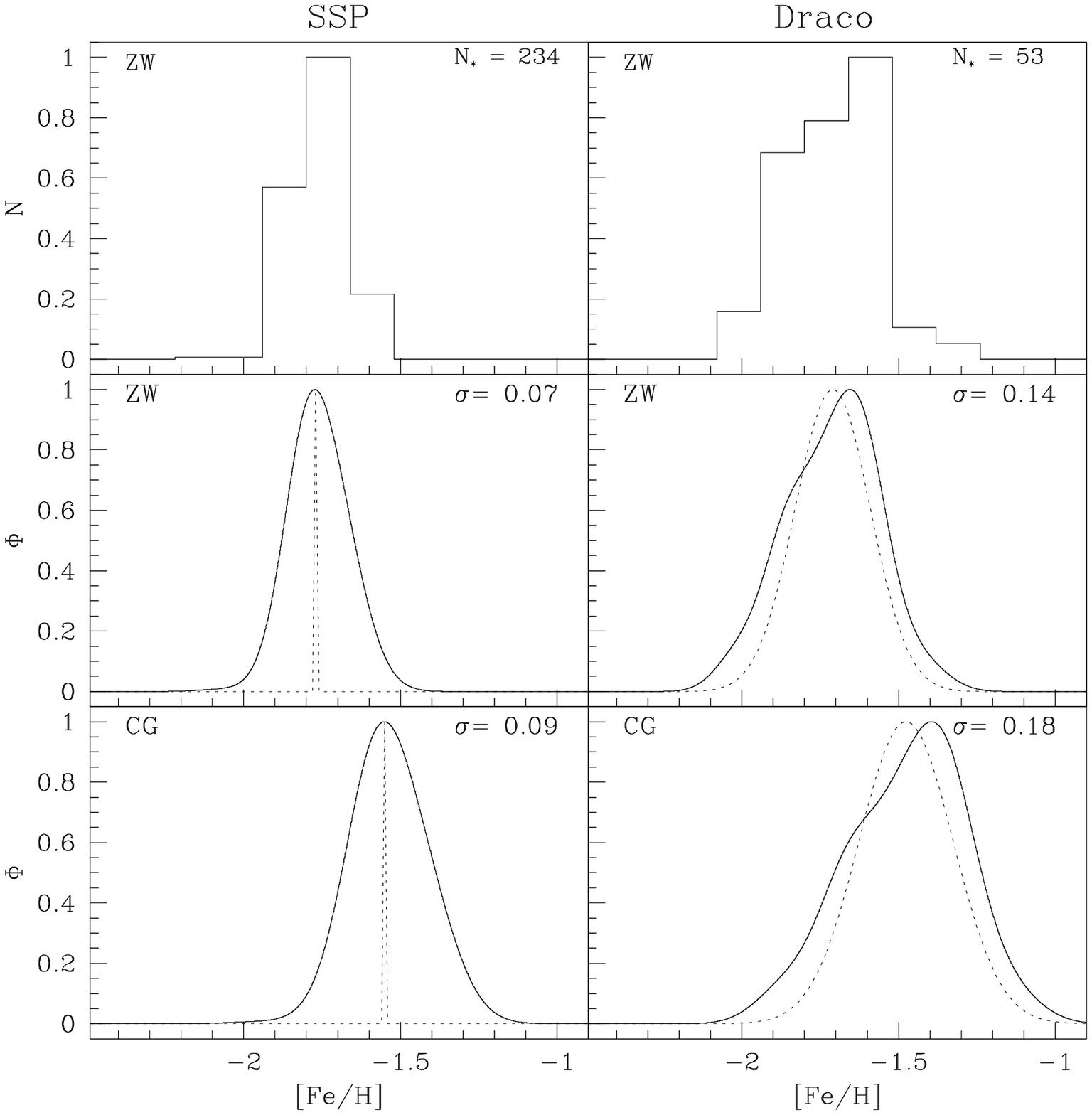}}
\caption{The photometric metallicity distribution of Draco (right panels)
are compared with the distribution obtained with the same technique from
the synthetic SSP we adopted for the artificial stars experiments
(left panels). Upper panels: odinary histograms in the ZW metallicity scale. 
Middle panels: generalized histograms in the ZW metallicity scale.
Lower panels: generalized histograms in the CG metallicity scale.
The  adopted smoothing lenght of the generalized histograms is equal to the 
measured $\sigma$ of the response function, i.e. the natural smoothing lenght.}
\end{figure*}

The  RGB color of old populations is mainly driven by the metal
content of the stars. Based on this principle, photometric Metallicity 
Distributions (MD) can be obtained by suitable interpolation between a grid of
metallicity templates, a technique that has been
widely applied in recent years \cite[see, e.g.][and references therein]{hol,
harris, ivo}. Here we obtained the MDs shown in Fig.~13 and Fig.~14 by
interpolating in color between the RGB ridge-lines of the adopted 
templates (see Fig.~12) for the stars in the magnitude 
range $-2.9\le M_I\le -3.9$.
While the accuracy of the single metallicity estimate is low, the
main characteristics of the distribution as a whole (as, for instance, its mean
and dispersion) are quite robust, being based on a large number of stars.
This technique is particularly appropriate in the present case since we know that
both galaxies are dominated by very old stellar populations
\citep{ken,grill,dol7,apadra,car02}.
However, the actual metallicity spread derived with this technique (a) is the
convolution of the intrinsic metallicity spread with the color spread due to
observational scatter and, (b) the impact of this latter factor {\em depends on
the mean metallicity}, since the same observational scatter should produce a
larger (apparent) metallicity spread in the low metallicity regime, where the
RGB ridge lines are steeper, with respect to the high-metallicity regime. 
In fact, $\Delta [Fe/H] / \Delta (V-I)$ decreases with increasing metallicity.
To deal with this problems we compare the observed photometric MDs
with those of the recovered artificial stars shown in Fig.~4,
that closely model the CMD of a SSP observed and reduced under the same
conditions as real stars.

\begin{figure*}
\figurenum{14}
\centerline{\psfig{figure=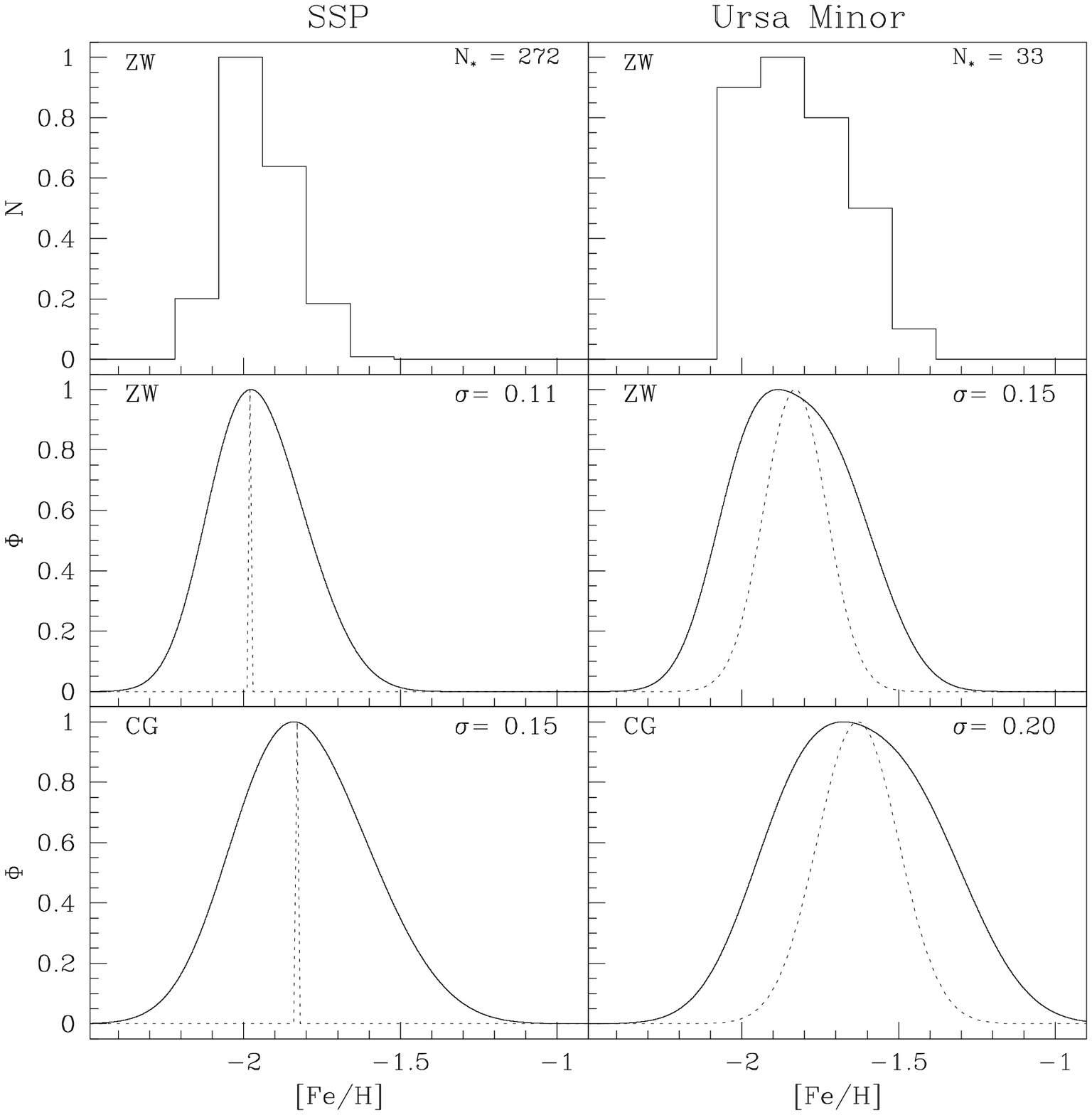}}
\caption{The photometric metallicity distribution of UMi (right panels)
are compared with the distribution obtained with the same technique from
the synthetic SSP we adopted for the artificial stars experiments.
The symbols are the same as in Fig.~13.
}
\end{figure*}

In Fig.~13 the MD of Draco (right panels) and the MD of
corresponding synthetic SSP (left panels; see
Fig.~5) are compared. All the MDs are normalized to their maximum value. 
The upper panels show the histograms in the ZW scale. The total
number of included stars is reported in the upper right corner.
The middle and lower panels show the
generalized histograms (continuous lines) in the ZW scale and in the CG scale, 
respectively \cite[see][and references therein]{hol,laird}. 
The measured standard deviation ($\sigma$) is reported in the 
upper right corner of each panel. The MD of the synthetic
SSP represents the response of the whole procedure to the pure observational
scatter, or in other words is the observed version of a {\em true} metallicity
distribution having the form of a Dirac delta function. The plots in the left
panels of Fig.~13 demonstrate that the observational scatter produce a sizeable
(spurious) metallicity spread. 
To obtain a sensible measure of the {\em true} intrinsic 
metallicity spread we have to deconvolve this ``response function'' from
the observed distributions shown in the right panels of Fig.~13. The dotted
lines are the gaussian distributions obtained by deconvolving the response 
function from a gaussian distribution having the same mean and standard
deviation of the observed distributions. The standard deviation of the dotted 
curves is a good estimate of the true {\em intrinsic} metallicity spread 
$\sigma_i$.

In the ZW scale the MD of Draco spans more than 0.6 dex.
The mean and median metallicities are $<[Fe/H]>=-1.7 \pm 0.1$, while the mode is
$[Fe/H]_{mod}=-1.6 \pm 0.1$. 
The intrinsic $1-\sigma$ scatter is $\sigma_i=0.13$ dex.
There is a marginal indication of a secondary peak at lower metallicty, 
but the statistical significance of this feature is low, 
thus we don't comment further on it.
These estimates are in good agreement with the results of the spectroscopic
analysis by \citet{len} and \citet{shet2}. The detailed comparison between
our photometric estimates and the study of \citet{shet2} is discussed in
\S5.2.1, below.
Our method is
insensitive to very metal poor stars ($[Fe/H]\le -2.5$), 
so they are obviously not represented in our metallicity distribution. 
However from Fig.~12 it may be evinced that the
number of possible RGB stars bluer (i.e. more metal poor) than the 
NGC~6341 template (at $[Fe/H]=-2.24$) should be small in both galaxy (see
\S5.2.1 for further details). 

In Fig.~14 the MDs of UMi and of its synthetic SSP are reported in the same way
as in Fig.~13. It can be appreciated that though the photometric
scatter is very similar for Draco and UMi (see Fig.~4) the response function has
a larger $\sigma$ in the case of UMi. This is due to larger sensitivity of
metallicity to color at the most metal poor regime described above. In the ZW
scale the MD of UMi spans a range nearly as wide as that of Draco.  
The mean and median metallicities are $<[Fe/H]>=-1.8 \pm 0.1$, while the mode is
$[Fe/H]_{mod}=-1.9 \pm 0.1$. The intrinsic scatter is $\sigma_i=0.10$ dex.
Also in this case the agreement with results of other studies 
\citep{shet2,apadra} and with the constraints obtained in \S5.1 is quite good.

From the comparison of the MDs of the two galaxies we can conclude that Draco is
slightly less metal deficient than UMi, in average, and it presents a slightly
larger metallicity spread. The parameters obtained from the photometric MDs 
(for Draco and UMi and for both metallicity scales) are summarized in 
Table~3. 

The MDs in the CG scale are remarkably similar to the MDs in the ZW scale 
described above, 
but, as expected, are shifted to higher metallicities and show a slightly
larger spread, thus the conclusions drawn above are unchanged. 
On the other hand, it
is important to remark that the comparison of the observed distributions with
chemical evolution models would lead to different conclusions depending on the
assumed metallicity scale. In particular, by adopting the CG scale, one should
conclude that the first stars in Draco and UMi formed from an ISM already
relatively enriched and reached a more advanced stage of chemical evolution
than suggested by the ZW scale. This kind of uncertainties in the
metallicity scales may have other dangerous drawbacks, hampering our
interpretation of astrophysical data in several cases 
\cite[see, e.g.][]{marcio}.

\subsubsection{Comparison with spectroscopic estimates}

A detailed comparison with the recent results by \citet{shet2} may be useful 
to check the consistency between high-resolution spectroscopic abundances and 
photometric metallicity estimates.
The stars observed by \citet{shet2} are plotted in Fig.~12 as large empty
symbols. Different symbols are adopted according to the metallicity estimates
by \citet{shet2}. Circles are stars with $[Fe/H]\le -2.10$, pentagons have
$-2.10 <[Fe/H]\le -1.8$, squares have $-1.8 <[Fe/H]\le -1.5$, and triangles
have $-1.5 <[Fe/H]\le -1.3$ (actually, there are only two stars in this latter
bin both having $[Fe/H]=-1.45$). In both panels of Fig.~12, the dashed 
line is the expected level of the TRGB with the adopted distance moduli (from
\S4.2) and the TRGB calibration by \citet{bfptip}. The dotted lines enclose the
full range of uncertainty in the expected magnitude of the TRGB, including
observational and calibration uncertainties.

First of all we note that all the considered stars have $I\ge I_{TRGB}$, to
within the uncertainties. This is compatible with the hypothesis that they are
first ascent RGB stars, as suggested by \citet{shet2}. Second, the metallicity
rank derived from spectroscopy is correctly reproduced by our photometry: stars
with higher spectroscopic metallicity estimates lie on redder RGB loci. Finally,
the zero-point of the spectroscopic scale is in good agreement (within the
uncertainties) with that
derived from our photometry, if the CG metallicity scale is adopted. 
Three out of
four of the stars in the more metal poor metallicity bin (circles) have 
$-2.36 \pm 0.09\le [Fe/H]\le -2.17 \pm 0.12$ and lie around the ridge line of
NGC~6341 ($[Fe/H]_{CG}=-2.16 \pm 0.10$). The two more metal rich stars 
(triangles, $[Fe/H] = -1.45 \pm 0.07$) are slightly bluer than the ridge line
of NGC~6205 ($[Fe/H]_{CG}=-1.39 \pm 0.10$). Hence both the metal rich and the
metal poor zero-points seems to agree quite well. The only exception is the star
\# 119 in Draco that is slightly redder than the ridge line of NGC~6341 and
that, according to \citet{shet2} has $[Fe/H]=-2.97 \pm 0.15$. We have no
suggestion to reconcile such a large discrepancy between color and spectroscopic
metallicity, except for trying to repeat both measures. 

If the ZW scale is adopted for the templates shown in Fig.~12 the
consistency in the metal poor zero-point is preserved
($[Fe/H]_{ZW}=-2.24 \pm 0.20$ for NGC~6341) but a serious problem emerges at the
metal rich end, where the two stars with $[Fe/H] = -1.45$ lie to the blue of
the $[Fe/H]=-1.65$ template and the $[Fe/H]=-1.4$ template (NGC~288) is more
than 0.1 mag redder than these stars. 
The problem with star \# 119 is not mitigated by the assumption of the ZW 
metallicity scale. Thus it appear that the adoption of the GC metallicity scale
provide a much higher degree of
consistensy between spectroscopic and photometric metallicity estimates, at
least in the present case. The above comparison may suggest that the
metallicities by \citet{shet2} are consistent with
the CG scale which is also based on high-resolution measures.

Finally, another possible source of uncertainty in the photometric metallicity
estimates is the small
difference in the average abundance of $\alpha$-elements between Dra and UMi 
and the template clusters \cite[see][and discussion therein]{shet2}.

\subsection{Populations gradients}

Radial population gradients are ubiquitous in dwarf galaxies 
\cite[see][and references therein]{grad,ivograd,hopp,tosi}. 
Almost in all cases
old/metal poor stars are found to be more abundant in the outer regions of dwarf
galaxies, while more metal rich and/or younger stars are preferentially 
found near center of galaxies \citep{grad}. 
Here we look for population gradients by using
the fraction of blue HB stars (i.e. those with $(V-I)<0.4$) in 
different concentric radial annuli. The small number of available HB stars
limits the spatial resolution to just three annuli. The results 
are shown in Fig.~15. In UMi (upper panel), 
the fraction of BHB stars
appears to rise slightly in the inner 4 arcmin. However the effect is at less
than $1 \sigma$, and the data are consistent with no population gradient, in
agreement with what found by \citet{car02}. The opposite effect is found in
Draco (lower panel): the BHB fraction drops by a factor 2 in the inner 4 arcmin,
i.e. the
classical gradient described above, with more metal rich and/or younger stars
clustered in the central region. The significance of the result is just marginal
($2 \sigma$), thus not in disagreement with the null detection reported by
\citet{apadra}.

\begin{figure*}
\figurenum{15}
\centerline{\psfig{figure=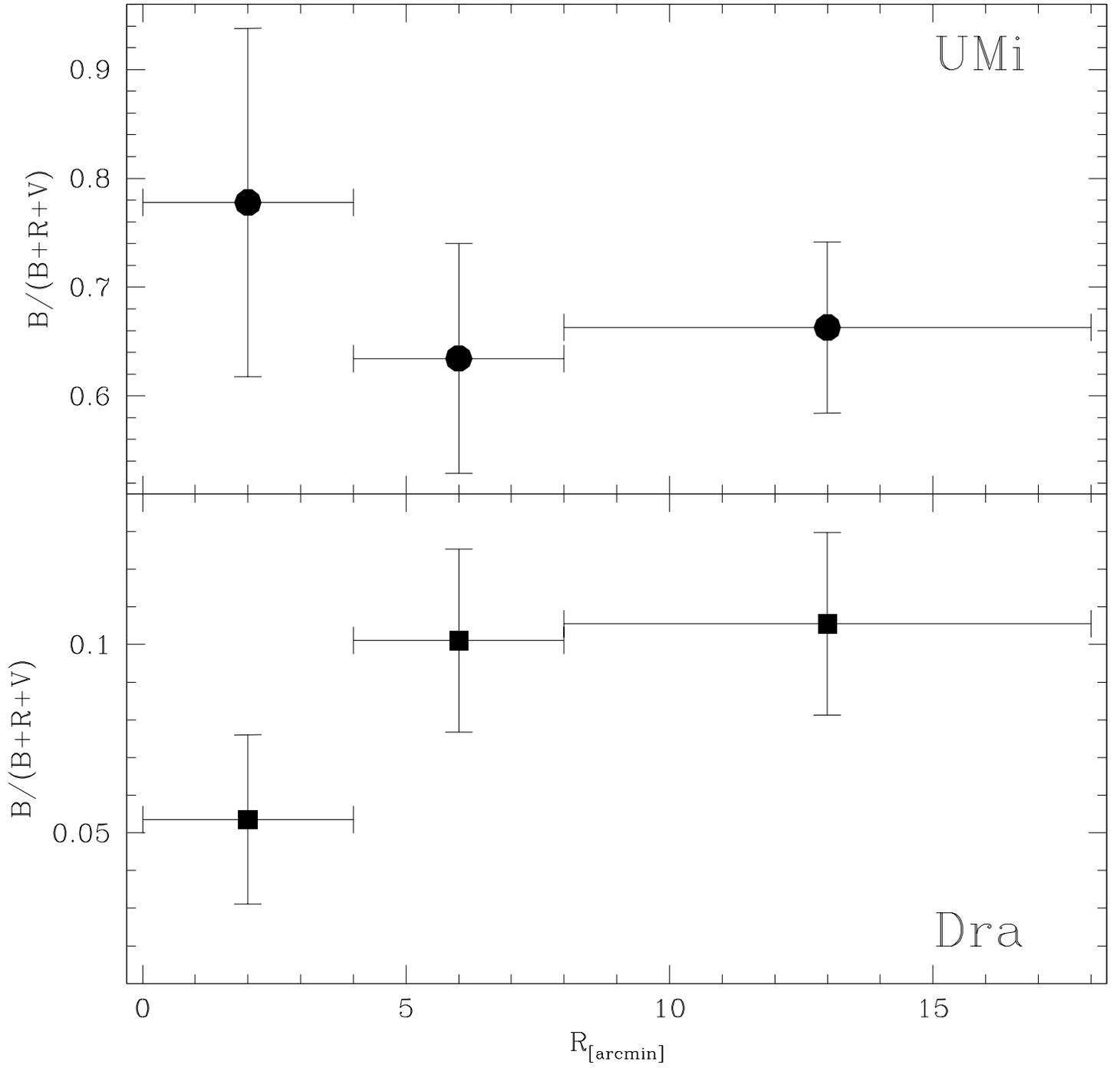}}
\caption{Radial population gradients in UMi (upper panel) and Draco (lower
panel), as measured by the fraction of BHB stars. The UMi data are consistent
with no gradient. The evidence for a population gradient in Draco is at the
2-$\sigma$ level.
}
\end{figure*}

\section{Structure}

We take advantage of our well defined CMDs to select likely members of UMi and
Draco with which reliable density contour maps can be obtained. The upper
panels of Fig.~16 show the adopted selections, while in the lower panels of the
same figure the isodensity contour maps of the selected members of UMi and 
Draco are shown. 
Our maps cover the inner regions of the galaxies,
within 1-2 core radii ($r_c$), and are obtained by computing the stellar 
density on an uniform grid with points spaced by 100 px. 
At each point of the grid the density
is computed over a circle of radius $r=3$ arcmin. This choice provides a 
suitable degree of smoothing and minimizes the effects of possible small empty
regions (f.i, saturated stars or dead CCD columns). The continuous lines are
isodensity contours starting from 1 $stars/arcmin^{2}$ and spaced by 0.5 
$stars/arcmin^{2}$. 
The innermost contour corresponds to 4.5 $stars/arcmin^{2}$ for UMi and to
6.0 $stars/arcmin^{2}$ for Draco. To give a term of comparison \citet{kle98} -
using a sample with similar limiting magnitude -
found a very similar value for the maximum density of UMi 
($\sim 4 ~stars/arcmin^{2}$) and estimated the
density of the background to be 0.31 $stars/arcmin^{2}$ at a distance of
$\sim 30$ arcmin from the center of UMi, much farther out than our
outermost isodense.  

The structure of the two galaxies, as shown in Fig.~16,
is in excellent agreement with previous results \citep{IH95,kle98,piatek,sdss}:
the ellipticity, position angle and the ratio between the central surface 
brightness are the same as reported in the review by \citet{m98}. 
Draco is more dense and has a much smoother profile with respect to UMi. 
Its isodensity contours are quite regular, showing just a marginal trend 
toward more boxy shapes in the outer parts. On the other hand the inner
contours of UMi appear quite structured and asymmetrical and deserve a deeper
analysis and discussion. 

The detection of  small scale clustering and 
asymmetric structures in the inner parts of UMi has been claimed by many 
different authors since the early '80 
\cite[see, e.g.][]{oa85,IH95,kle98,db99,db01,esk}, but a clear-cut demonstration
of the statistical significance of such structures is still lacking.
In Fig.~16 the isodensity contours up to 3.0 $stars/arcmin^{2}$ appear quite
elongated but similar in shape and symmetrical. On the other hand the peak of
density ($X\simeq 1.8$;$Y\simeq 3$) is clearly {\em off-centered} with respect 
to the center of symmetry of the outer contours ($X\simeq 4.3$;$Y\simeq 5.3$). 
We note that the density contrast between the density peak and the last
symmetric contour is larger than 4 $\sigma$, i.e. quite significant.
It is also remarkable that the off-centered main density peak displays a much
rounder shape, with respect to the rest of the galaxy.
On the
other hand, the dotted contour (corresponding to a density of 3.2 $stars/arcmin^{2}$)
is reported to show that the substructures in the NE region of the galaxy that
are clearly visible in the maps by \citet{IH95} and \citet{kle98} are present 
also in our
sample but have a low density contrast with respect to the last symmetric
contour, thus low statistical significance, as concluded also by \citet{kle98}.

\begin{figure*}
\figurenum{16}
\centerline{\psfig{figure=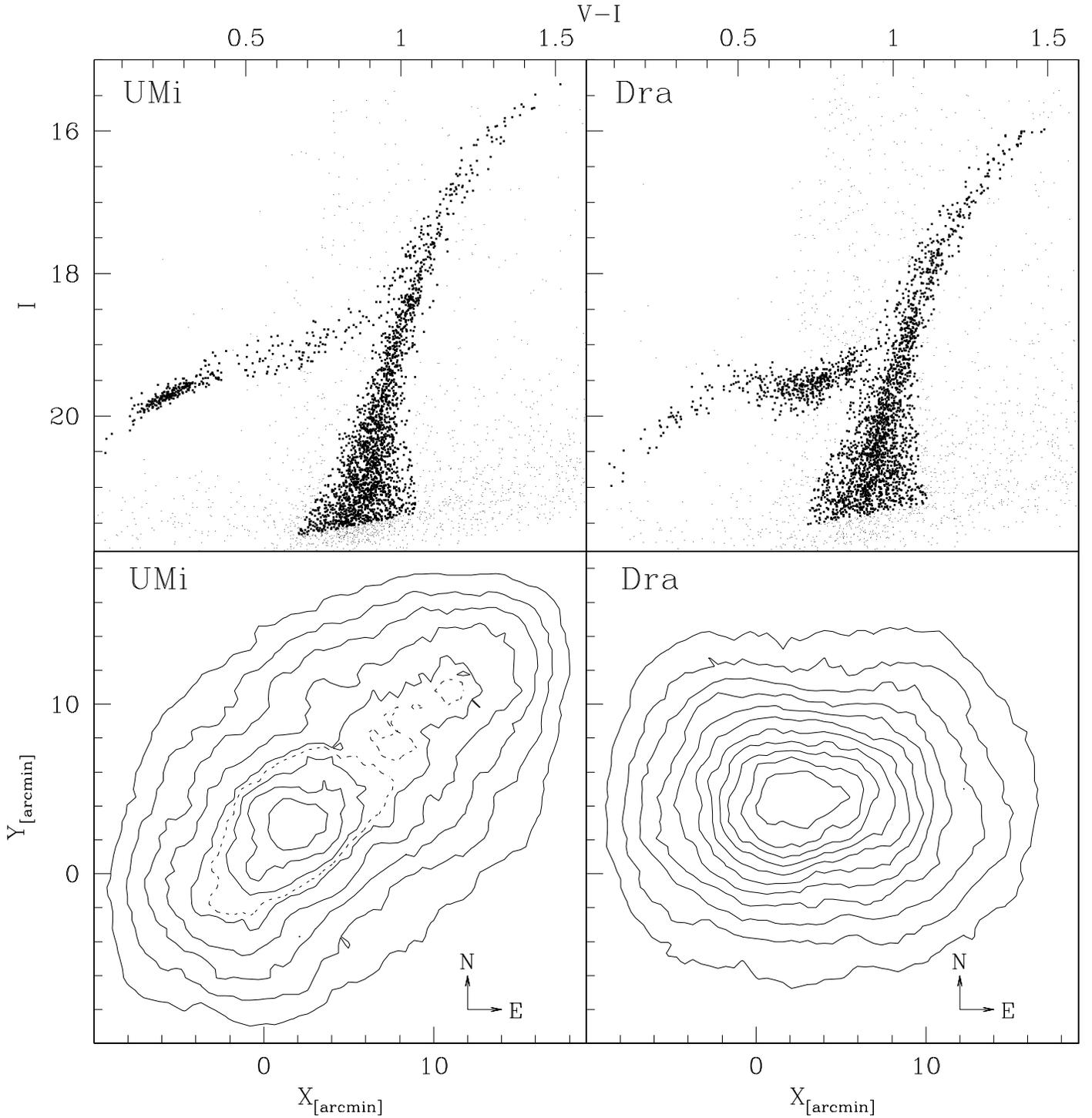}}
\caption{Upper panels: CMDs of the two galaxies. The stars selected as members
(bold face dots) are used to derive the density contour maps shown in the lower
panels. The outermost contour corresponds to a density of 1 star arcmin$^{-2}$; 
the contours are plotted in steps of 0.5 stars arcmin$^{-2}$ 
(i.e. 1.0, 1.5, 2.0, ..., etc.). 
The innermost contour corresponds to a density of 4.5 stars 
arcmin$^{-2}$ for UMi, and 6.0 stars arcmin$^{-2}$ for Draco. 
The dotted contour
in the map of UMi corresponds to a density of 3.2 stars arcmin$^{-2}$. It has
been reported to show the low contrast structures in the north-eastern region 
of the galaxy that have already been noted by \citet{IH95}. The dotted contour
shows also that the symmetry seen in the outer isodense curves breaks just above
the density of 3 arcmin$^{-2}$.
The density is estimated over a 100
pixel spaced grid. At each point of the grid the density is estimated in a
circle with radius $= 3$ arcmin. 
}
\end{figure*}

It is important to remark that the only other spheroidal galaxy of the Local
Group that has a clearly off-centered density peak is the Sagittarius dSph,
whose structure is greatly strained by the Galactic tidal field
\cite[see][]{ibata,stream}. Independent evidence of the effects of the 
Galactic tidal fields on UMi has been reported by \citet{umitail}.

\citet{kle98} found significant
asymmetry in the distribution of stars along the major axis. Here we concentrate
on the clustering properties of UMi stars. We define $d_n$ as the distance of a
given star to its n$-th$ nearest neighbour. In the following we will show the
results obtained with $d_{200}$, i.e. the distance to the 200$-th$ nearest
neighbour, but we remark that the results are preserved for a very wide range 
of $n$ ($50\le n \le 500$). In the upper panel of Fig.~17 the distributions of  
$d_{200}$ are shown for three template samples having the same dimensions of the
observed ones ($\sim 2500$ member stars per galaxy). The dotted curve is the
$d_{200}$ distribution of a sample randomly drawn from a uniform distribution.
The continuous line and dashed lines are the $d_{200}$ distributions of
samples drawn from an elliptical Gaussian
distribution having $\sigma_x = r_c(UMi)$ and $\sigma_x/\sigma_y = (a/b)_{UMi}$,
and $\sigma_x = r_c(Dra)$ and $\sigma_x/\sigma_y = (a/b)_{Dra}$, respectively,
where {\em a} and {\em b} are the semi-major 
and semi-minor axis of the galaxies, as listed by \citet{m98}.
It can be easily appreciated that (1) more concentrated (denser) systems 
have $d_{200}$ distributions peaked at shorter lenghts and (2) independently of
the assumed density profile, the $d_{200}$ distributions of symmetric and smooth
systems show a single major peak, marking the characteristic clustering scale,
followed by {\em a very sharp cut-off}. In these systems there are no stars 
having $d_{200}$ shorter than the cut-off value. 

In the middle panel of Fig.~17 the $d_{200}$ distribution of Draco is presented.
The observed distribution is very well
reproduced by its template: the characteristic clustering scale is
$\simeq 500$ px $\simeq 2.3\arcmin$ and no star with clustering scale shorter 
than the cut-off 
($\simeq 480$ px) is observed. On the other hand, the $d_{200}$ distribution
of UMi (shown in the lower panel of Fig.~17) is quite different from that of
Draco and from the smooth/symmetric templates. The major peak and the cut-off
are well reproduced by the corresponding template but {\em there is a
clear secondary peak at shorter scale lenghts with respect to the main cut-off}.
The presence of this secondary peak shows that there are {\em two characteristic
scales of clustering} in UMi, a longer one ($\simeq 700$ px $\simeq 3.2\arcmin$)
associated with the general properties of the system and a shorter one 
($\simeq 600$ px $\simeq 2.7\arcmin$) associated with the more
compact structures corresponding, to the off-centered density 
peak.

\begin{figure*}
\figurenum{17}
\centerline{\psfig{figure=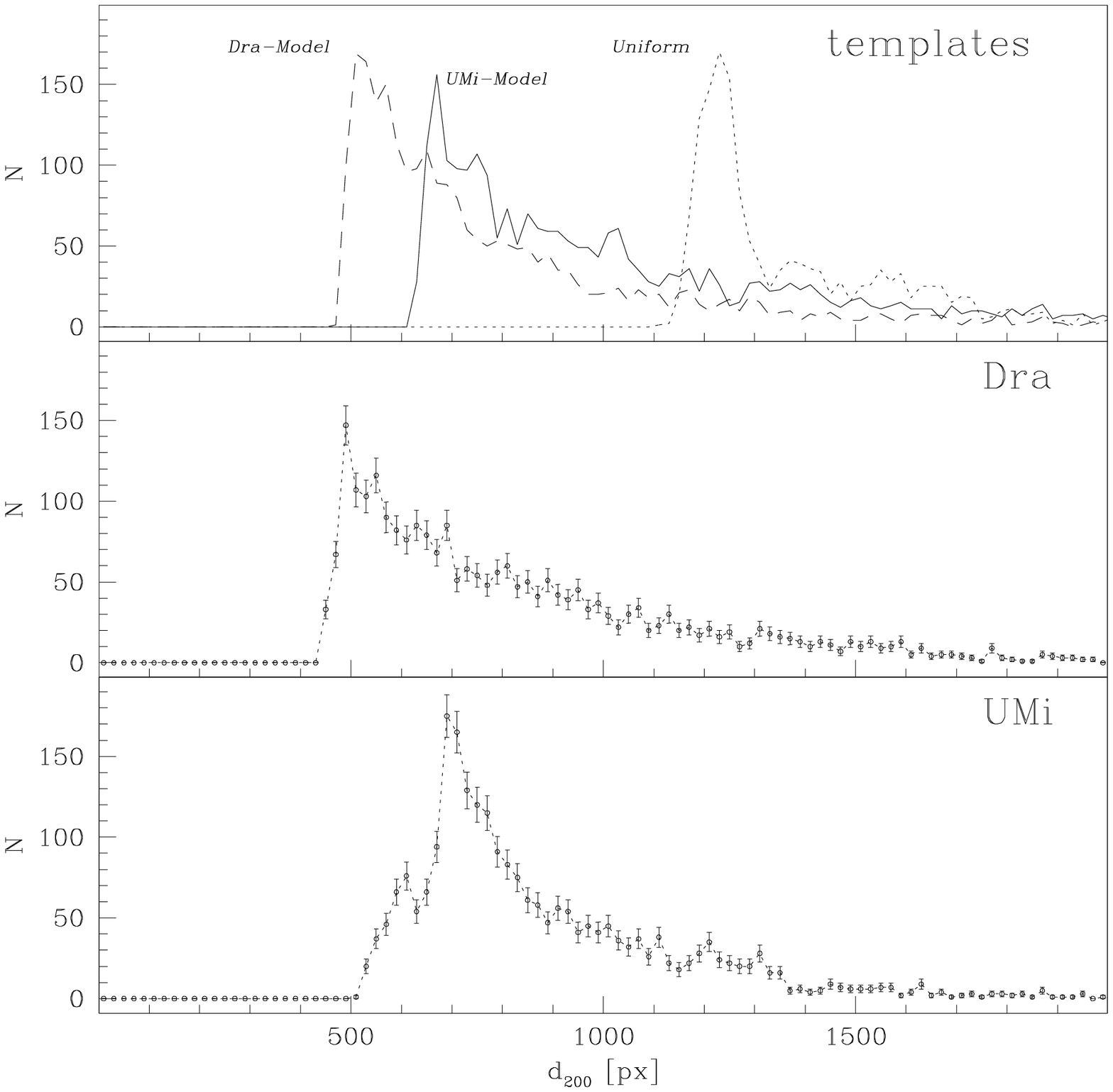}}
\caption{Distributions of the distances to the 200-$th$ nearest neighbour for:
(upper panel) a uniform template and two templates with the same characteristic
radius and axis ratio as UMi and Dra (see text); (middle panel): the selected
candidate members of Draco; (lower panel) the selected candidate members of UMi.
}
\end{figure*}

Can the observed $d_{200}$ distribution of UMi arise by chance (i.e. small
number fluctuations) from an intrinsically smooth and unstructured system ?
To answer this question we extracted 1000 random samples from the model
assumed for the UMi template shown in the upper panel of Fig.~17 and for 
each of them we computed
the fraction of stars having $d_{200}$ lower than the observed cut-off 
($F(d_{200}<d_{200}^{cut-off}$, hereafter $F_d$, for brevity), 
fixed at $d_{200}=630$ px. 
The distribution of the $F_d$ of the simulated samples
is shown in Fig.~18. The observed fraction ($F_d=0.12$) is indicated by a dotted
line. Fig.~18 shows that (a) in no case a sample having $F_d$ equal or 
greater than the observed value has been extracted from the unstructured model, 
(b) the highest $F_d$ value reached by a simulation ($F_d\simeq0.08$) is
significantly lower than the observed one, and (c) 95 \% of the simulated
samples have $F_d\le0.04$. We conclude that {\em the observed presence of two
characteristic clustering scales is a real and statistically significant 
property of UMi}. 

\begin{figure*}
\figurenum{18}
\centerline{\psfig{figure=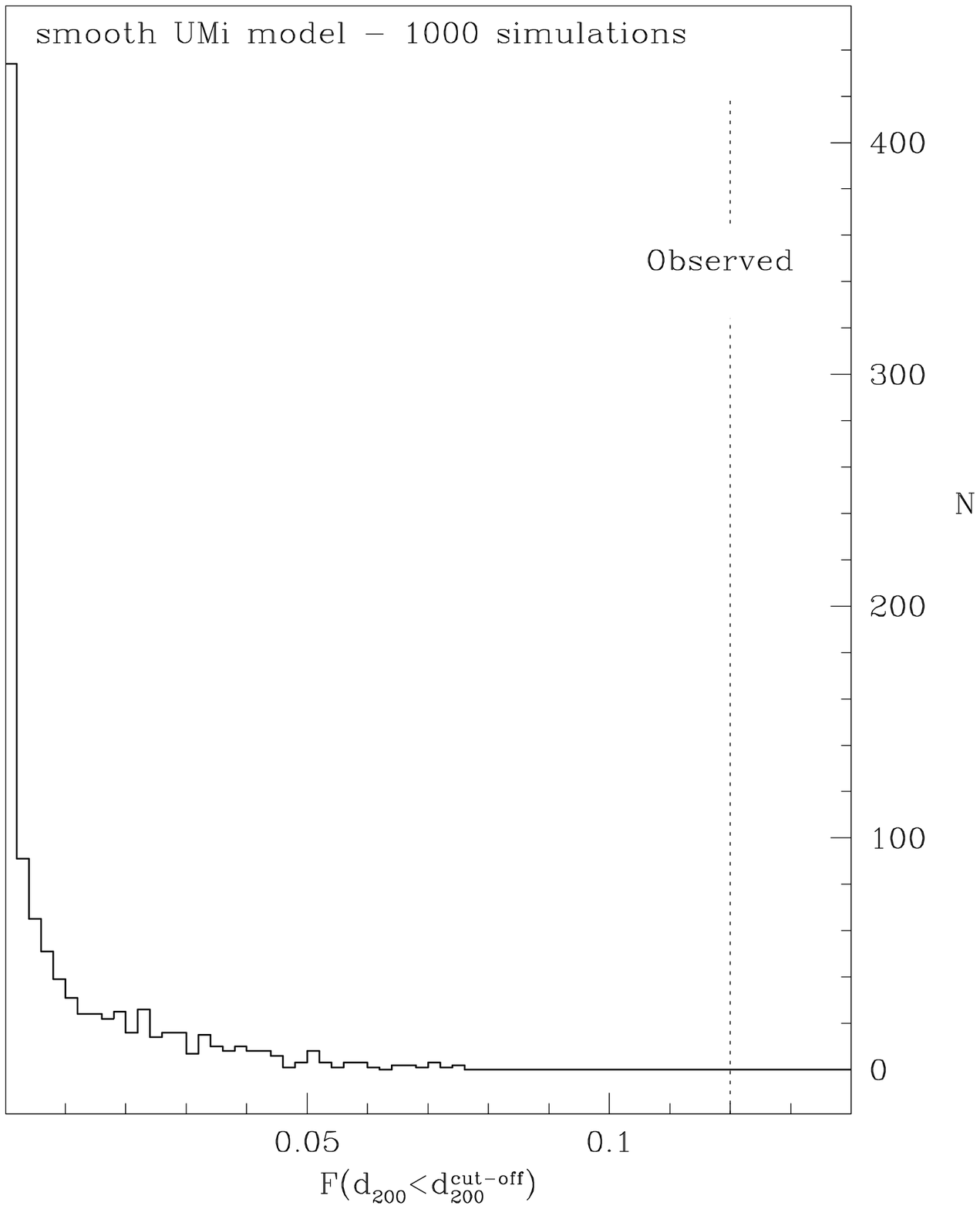}}
\caption{Frequency distribution of the stars with $d_{200} <$ cut-off value 
[$F(d_{200}<d_{200}^{cut-off}$] drawn from 1000 random samplesequivalent to the
unstructured model of UMi shown in the upper panel of Fig.~17.
The dotted line marks the value of 
$F(d_{200}<d_{200}^{cut-off}$ actually observed in UMi, $\sim 1.5$ times larger
than the largest value obtained in these simulations. This test shows that the
second peak in the $d_{200}$ distribution of UMi cannot arise by chance from a
smooth and symmetrical model.
}
\end{figure*}

Finally, we find that the spread in magnitude of the blue HBs of Draco and UMi 
are small and compatible with the observational scatter at that magnitude. 
Thus we can't see any sign of the significant elongation along the line of 
sight that was required in the model by \citet{kk98} to explane the large 
velocity dispersion observed in dSph galaxies without the need of massive dark 
matter halos \cite[see also][]{apadra}.

\section{Discussion}

The distance estimates to Draco and UMi derived in \S4 are tied to the distance
scale introduced by F99  and compatible with the TRGB scale introduced 
by \citet{bfptip}.
The F99 scale is fully consistent with the scale based on the revised
Hipparcos parallaxes by \citet{car00}. F99 showed that their distance moduli of
globular clusters are tipically $\sim 0.2$ mag larger than those tied to
traditional $M_V(RR) - [Fe/H]$ calibrating relation. This is probably the main
reason why we find distance moduli for UMi and Draco that are larger by $0.2 -
0.3$ mag than what generally found in literature 
\cite[see, e.g.][and references therein]{m98}. We note also that, independently
of the distance scale,  our procedure
of HB matching leads to results that are in excellent agreement with works 
based on
the RR Lyrae \cite[e.g., N88][]{nemdra} while a sensible mismatch is noticed
with respect to the Main Sequence fitting technique adopted by \citet{ken},
using HST data. We argue that the problems with the absolute photometric
calibration of HST-WFPC2 data may be at the origin of the observed differences.

The comparison with the recent distance estimates by \citet{apadra} (for Draco)
and \citet{car02} (for UMi) is of particular interest.
\citet{apadra} derived a distance modulus to Draco by (1) estimating  
$V_{RR}$ from the observed HB and (2) by adopting the classical 
$M_V(RR)-[Fe/H]$ 
relation by \citet{lee}. They obtain $V_{RR}=20.14 \pm 0.12$, compatible with 
our estimate within the uncertainties, but their final distance modulus 
($(m-M)_0=19.5 \pm 0.2$) is 0.34 mag smaller than ours. However, if the above
quoted systematic difference of $\sim 0.2$ mag between the F99 distance scale 
and the scale by \citet{lee} is taken into account the consistency between the 
two estimates is clearly recovered. On the other hand, \citet{car02} derived
the mean level of RR Lyrae in UMi by a comparison with template globular cluster
of similar metallicity, i.e. the same method adopted here. Moreover, the 
distance moduli of the template clusters were taken from \citet{reid} who
derived them from Main Sequence fitting to Hipparcos subdwarfs. With this
approach \citet{car02} obtained $V_RR = 19.84\pm 0.07$ and 
$(m-M)_0=19.40 \pm 0.10$ in {\em excellent} agreement with our results.

Independently of the above considerations, it is important to remark that the
actual uncertainty on state-of-the-art distance estimates to Draco and UMi
remains $\sim \pm 25$ \%, in spite of all efforts. It is clear that classical
distance indicators are unefficient in this context and valid
alternatives  are certainly needed. The search for
detached double-lined eclipsing binaries (DD-EB) is now possible with new 
generation instruments and may be very rewarding, given the high binary 
fraction that has been reported for these two galaxies \citep{olsbin}. 

\subsection{Evolutionary history}

The metallicity distributions presented here confirm (with larger samples, 
with respect to existing spectroscopic surveys)
that the stellar populations of both galaxies have a 
sizeable spread in metal content. 
Since they are dominated by very old stars, we shall conclude that 
significant self-enrichment took place in these systems in a quite
short time scale (i.e., few Gyr) at very early epochs. The same conclusions have
been previously reached for Sculptor \citep{sculp} and Sextans \citep{bfpsex}.
\citet{shet2} have shown that some of the stars of UMi, Dra and Sex show
abundance patterns significantly different from those typical of Galactic halo
stars. In particular it has been suggested \citep{ikuta} that the transition 
from an $\alpha$
-enhanced regime to nearly solar $[\alpha/Fe]$ ratios occurred at a much lower
[Fe/H] value than in our own Galaxy. This evidence led  \citet{shet2} to 
conclude that it is unlikely that systems like UMi and Draco gave a
significant contribution to the assembly of the Galactic halo. We remark here
that this conclusion is correct only for satellites that would have been
accreted by the Milky Way in an advanced stage of their chemical evolution.
A very early accretion (i.e., within
$\sim 1$ Gyr from the onset of star formation) or simply a very early stripping
of the gaseous component by the Galaxy wouldn't leave any peculiar chemical
signature in a Galactic halo largely composed by building blocks like Draco, UMi
or Sextans (as they were in their first Gyr of ``stellar'' life). 

Our results show also that, while the average metallicity of the two galaxy is
similar, the difference in the MDs is significant. The MD of UMi peaks at
$[Fe/H]\simeq -1.9$ and barely reaches $[Fe/H]\simeq -1.6$ while that of Draco
peaks at $[Fe/H]\simeq -1.6$ and reaches $[Fe/H]\simeq -1.3$. This fact has to
be taken into account, in particular in the interpretation of the great
difference in HB morphology between the two galaxies. Dra and UMi {\em are not}
a good ``second parameter pair'' \cite[see][and references therein]{mb01}: a
sizeable part of the HB morphology difference may be due to the differences in
the actual MD. The large population of binaries (and Blue Stragglers) hosted by 
the galaxies \citep{olsbin} may also be have some impact on the observed HB
morphology, especially in the case of Draco  
\cite[see][and references therein]{ffp92,mb01,mb02a,carney}.

\subsection{A striking structural difference: not so twins after all?}

Draco and UMi have similar distance from the center of the Galaxy, similar
luminous mass and similar star formation histories. Both are thought to have
high M/L ratios and probably have also similar orbits, since they have similar
radial velocities and (possibly) proper motions \cite[see][]{m98,SH94,cud}. 
In spite of that,
while Draco shows a very symmetrical and smooth profile out to a very large
distance from the center \citep{sdss,piatek}, we have demonstrated that 
{\em even the innermost regions} of UMi are strongly structured and
asymmetrical. The presence
of a massive and extended CDM halo should have inhibited the formation of
any significant substructure in the stellar component of UMi. 
The star formation
activity cannot be responsible of  the disturbed structure of this galaxy since
significant star formation episodes have ceased many Gyr ago. Furthermore we
accurately checked for possible differences in the stellar populations between 
the off-centered (and nearly round) density peak of UMi and its outer regions 
and we found none.  

\citet{omega} showed that a small self-gravitating stellar system is embedded
into the main body of the peculiar globular cluster $\omega$ Cen. This stellar
system has likely managed to preserve its individuality over many Gyr, while
orbiting into the stellar halo of the cluster. A similar possibility could be
envisaged also to explain the origin of the peculiar density peak of UMi. The
process of hyerarchical merging is expected to be scale-free, thus the records
of the formation of a galaxy like UMi from smaller fragments may be still
detectable today \citep[see][]{kroupa}, possibly favoured by the fact that the
putative fragment we are considering is much denser than the stellar medium it 
is embedded in. 
The main argument against this hypothesis is the already reported
observational evidence that the stellar population in the density peak does not
differ from the average UMi population 
\cite[as opposite to the $\omega$ Cen case, see][and references therein]{omega}. 
Large high-resolution spectroscopic
surveys may possibly provide more stringent indications on the viability of this
scenario.

While we are writing, a new preprint was posted 
\citep{palma} in which our results about the structure of UMi are fully
confirmed\footnote{However the clustering properties of UMi stars are not
considered in that work.}. 
Furthermore \citet{palma} were able to follow the structure of UMi
out to a distance of $\sim 1.7$ deg from the center and demonstrated that 
(1) the galaxy is
elongated along the direction of its proper motion vector and (2) the outer
isodensity contours of UMi have the typical S-shape of tidally disturbed 
stellar systems. 
If it will be spectroscopically confirmed that the stars in the
S-shaped structures are unbound, or the signature of apparent rotation will be
found, this would directly disprove a specific prediction of the CDM simulations
\citep{haya}. In any case, the strongly disturbed structure of UMi provides a
severe challenge to standard CDM scenarios, above all if the comparison with
Draco is considered\footnote{In fact, the presently available observations of
Draco are very well fitted by the predictions of CDM models \citep{haya}}. 
Moreover, \citet{palma} showed that previous estimates of
the total luminosity of UMi had missed a significant amount of
light that is found at large distances from the center of the galaxy. Their
estimate of total luminosity is 2.7 times larger than previous ones. Taking this
into account and considering the (possible) effects of anisotropic velocity 
dispersions they were able to reduce the mass to light ratio of UMi down to
$M/L\simeq 16$. If we include in this computation the effects of our larger
distance modulus 
\cite[and assuming $M_{tot}(UMi)=2.3\times 10^7 M_{\odot}$, after][]{m98} 
we obtain $M/L\simeq 7$, just a factor $\sim$ 4 larger  than the typical M/L of 
ordinary globular clusters.

From the above discussion it appear that further observational and theoretical
efforts are still needed to fully understand the nature of UMi and Draco. It is
quite possible that we have some fundamental lesson to learn  from these faint 
and unassuming stellar systems. 

\acknowledgments
This research has been  partially supported by  the
{\it Agenzia Spaziale Italiana} (ASI)  and by  the 
{\it Ministero dell'Universit\`a e della Ricerca Scientifica
 e Tecnologica} (MURST).
E.P. acknowledges the support of the ESO Studentship Programme.
Carla Cacciari is warmly thanked for a critical reading of the original
manuscript. We thank M. Shetrone for providing the coordinates of his
target stars.
We are grateful to the staff of 
the Telescopio Nazionale Galileo (TNG)
for the warm hospitality and the professional assistance.
Part of the data analysis has been performed using 
software developed by P. Montegriffo at the Osservatorio Astronomico
di Bologna. This research has made use of NASA's
Astrophysics Data System Abstract Service. 


\clearpage


\begin{deluxetable}{lcccccccccc}
\tablecolumns{11}
\tablewidth{0pc}
\tablecaption{Photometry and positions of stars in Ursa Minor.}
\tablehead{
\colhead{ID} & \colhead{V} &\colhead{$\epsilon_V$}  & \colhead{I} & 
\colhead{$\epsilon_I$} & \colhead{$X_{[px]}$} & \colhead{$Y_{[px]}$} &
\colhead{$\alpha_{2000}$}&\colhead{$\delta_{2000}$}  &\colhead{Var. type} 
&\colhead{Var. n.}\\}
\tabletypesize{\footnotesize}
\startdata
 10018 & 16.858 & 0.005 & 15.342 & 0.005 &  246.08 &  365.12 &  15$^h$ 08$^m$ 27.2$^s$  &  67$\degr$ 10$\arcmin$ 07.7$\arcsec$  & 0 & 0\\
 10019 & 17.018 & 0.006 & 15.628 & 0.006 &  399.38 &  462.34 &  15$^h$ 08$^m$ 34.4$^s$  &  67$\degr$ 10$\arcmin$ 34.2$\arcsec$  & 0 & 0\\
 10020 & 16.542 & 0.004 & 15.717 & 0.005 &  754.09 &  801.17 &  15$^h$ 08$^m$ 51.2$^s$  &  67$\degr$ 12$\arcmin$ 07.0$\arcsec$  & 0 & 0\\
 10021 & 16.579 & 0.005 & 15.620 & 0.008 & 1959.44 &  817.81 &  15$^h$ 09$^m$ 48.4$^s$  &  67$\degr$ 12$\arcmin$ 09.3$\arcsec$  & 0 & 0\\
 10023 & 16.928 & 0.004 & 15.679 & 0.006 &  838.95 & 1488.32 &  15$^h$ 08$^m$ 55.4$^s$  &  67$\degr$ 15$\arcmin$ 15.9$\arcsec$  & 0 & 0\\
 10024 & 16.458 & 0.005 & 15.569 & 0.008 &  873.74 & 1918.16 &  15$^h$ 08$^m$ 57.2$^s$  &  67$\degr$ 17$\arcmin$ 14.2$\arcsec$  & 0 & 0\\
 10025 & 17.714 & 0.005 & 16.489 & 0.005 &  548.78 &   65.36 &  15$^h$ 08$^m$ 41.5$^s$  &  67$\degr$ 08$\arcmin$ 44.9$\arcsec$  & 0 & 0\\
 10026 & 17.459 & 0.006 & 16.467 & 0.009 &  102.13 &  160.05 &  15$^h$ 08$^m$ 20.4$^s$  &  67$\degr$ 09$\arcmin$ 11.5$\arcsec$  & 0 & 0\\
 10028 & 18.090 & 0.006 & 15.754 & 0.007 & 1399.35 &  279.70 &  15$^h$ 09$^m$ 21.7$^s$  &  67$\degr$ 09$\arcmin$ 42.5$\arcsec$  & 0 & 0\\
\nodata&\nodata&\nodata&\nodata&\nodata&\nodata&\nodata&\nodata&\nodata&\nodata&\nodata\\
\enddata
\tablecomments{A sample of the photometric catalog. The complete catalog is
available in ASCII format in the electronic edition of the paper. The variable
stars are numbered after the nomenclature by N88 (column 11) and classified
according the same authors (column 10: 1 = RRab; 2 = RRc; 3 = Anomalous Cepheids;
4 = unclassified variables).}
\end{deluxetable}

\begin{deluxetable}{lcccccccccc}
\tablecolumns{11}
\tablewidth{0pc}
\tablecaption{Photometry and positions of stars in Draco.}
\tablehead{
\colhead{ID} & \colhead{V} &\colhead{$\epsilon_V$}  & \colhead{I} & 
\colhead{$\epsilon_I$} & \colhead{$X_{[px]}$} & \colhead{$Y_{[px]}$} &
\colhead{$\alpha_{2000}$}&\colhead{$\delta_{2000}$}  &\colhead{Var. type} 
&\colhead{Var. n.}\\}
\tabletypesize{\footnotesize}
\startdata
 10069 & 16.744 & 0.004 & 15.963 & 0.003 & 1295.30 &  338.62 & 17$^h$ 20$^m$ 34.3$^s$  & 57$\degr$ 51$\arcmin$ 56.4$\arcsec$  & 0 & 0\\
 10085 & 17.510 & 0.004 & 16.742 & 0.003 & 1375.32 &  999.17 & 17$^h$ 20$^m$ 37.2$^s$  & 57$\degr$ 54$\arcmin$ 58.2$\arcsec$  & 0 & 0\\
 10086 & 17.478 & 0.003 & 16.802 & 0.003 & 1145.29 & 1051.62 & 17$^h$ 20$^m$ 29.3$^s$  & 57$\degr$ 55$\arcmin$ 13.1$\arcsec$  & 0 & 0\\
 10090 & 16.989 & 0.004 & 15.710 & 0.002 &  505.79 & 1158.47 & 17$^h$ 20$^m$ 07.2$^s$  & 57$\degr$ 55$\arcmin$ 43.5$\arcsec$  & 0 & 0\\
 10091 & 17.313 & 0.003 & 15.744 & 0.003 & 1828.06 & 1218.93 & 17$^h$ 20$^m$ 52.9$^s$  & 57$\degr$ 55$\arcmin$ 57.8$\arcsec$  & 0 & 0\\
 10097 & 17.112 & 0.003 & 15.816 & 0.002 &  463.32 & 1303.23 & 17$^h$ 20$^m$ 05.7$^s$  & 57$\degr$ 56$\arcmin$ 23.4$\arcsec$  & 0 & 0\\
 10100 & 17.212 & 0.004 & 15.759 & 0.002 &  262.30 & 1511.18 & 17$^h$ 19$^m$ 58.8$^s$  & 57$\degr$ 57$\arcmin$ 20.9$\arcsec$  & 0 & 0\\
 10109 & 17.193 & 0.003 & 16.272 & 0.002 & 1456.76 & 1651.69 & 17$^h$ 20$^m$ 40.2$^s$  & 57$\degr$ 57$\arcmin$ 57.7$\arcsec$  & 0 & 0\\
 10110 & 16.948 & 0.003 & 15.746 & 0.002 &  270.94 & 1652.71 & 17$^h$ 19$^m$ 59.2$^s$  & 57$\degr$ 57$\arcmin$ 59.8$\arcsec$  & 0 & 0\\
\nodata&\nodata&\nodata&\nodata&\nodata&\nodata&\nodata&\nodata&\nodata&\nodata&\nodata\\
\enddata
\tablecomments{A sample of the photometric catalog. The complete catalog is
available in ASCII format in the electronic edition of the paper. The variable
stars are numbered after the nomenclature by \citet{baade} (column 11). 
In column 10 only the known Anomalous Cepheid V141 is flagged with
Var. type = 3.}
\end{deluxetable}

\clearpage

\begin{deluxetable}{lccc}
\tablecolumns{4}
\tablewidth{0pc}
\tablecaption{Physical parameters for Ursa Minor and Draco.}
\tablehead{
\colhead{} & \colhead{UMi} &\colhead{Draco}  & \colhead{ref.} \\ }
\startdata
E(B-V)        & 0.03 & 0.03  & M98	\\
$V_{RR}$      & 19.86 $\pm$ 0.09 & 20.28 $\pm$ 0.10 &	\\
$V_{ZAHB}$    & 20.02 $\pm$ 0.09 & 20.44 $\pm$ 0.10 &	\\
$V^{bump}$    & 19.40 $\pm$ 0.06 & \nodata &	\\
$(m-M)_0$     & 19.41 $\pm$ 0.12 & 19.84 $\pm$ 0.14 &	\\
$r_c $         & $17.9\arcmin$ & $7.7\arcmin$ & P02, O01	\\
$r_t $         & $77.9\arcmin$ & $40.1\arcmin$ & P02, O01\\
$e=(1-b/a)$   & 0.80 & 0.29 & IH95	\\
$M_V$         & $-10.3\pm 0.4$ & $-9.0\pm 0.3$ & M98, P02, this work	\\
$\mu_V(0)$[mag/$\sq \arcsec$]    & $25.5\pm 0.5$ & $25.3\pm 0.5$ & M98	\\
$<[Fe/H]>$ (ZW)  & $-1.8\pm 0.1$  & $-1.7\pm 0.1$ &	\\
$[Fe/H]_{mod}$ (ZW) & $-1.9\pm 0.1$ & $-1.6\pm 0.1$ &	\\
$<[Fe/H]>$ (CG)  & $-1.6\pm 0.1$  & $-1.5\pm 0.1$ &	\\
$[Fe/H]_{mod}$ (CG)      & $-1.7\pm 0.1$ & $-1.4\pm 0.1$ &	\\
$\sigma_i^{[Fe/H]} (ZW) $  & 0.10 & 0.13 &	\\
$\sigma_i^{[Fe/H]} (CG) $  & 0.12 & 0.15 &	\\
\enddata
\tablecomments{A summary of relevant observational parameters for UMi and Dra.
If no explicit reference is indicated, the estimated obtained in the present
analysis is reported. The reported $M_V$ has been obtained adopting our distance
moduli. Legenda of the previously undefined acronyms: P02 = \citet{palma};
O01 = \citet{sdss}; IH95 = \citet{IH95}. The reported values of $M_V$ have been
obtained from the apparent total magnitudes listed by M98 (corrected according
to the results by P02, in the case of UMi) and the distance moduli obtained in
the present analysis.}
\end{deluxetable}

\clearpage

\end{document}